\documentclass[aps,prd,superscriptaddress,twoside,twocolumn,nofootinbib,showpacs,
floatfix]{revtex4-2}
\usepackage{amsmath,amssymb}
\usepackage{graphicx,bm}
\usepackage{slashed,color}
\usepackage[colorlinks=true,pdfstartview=FitV,bookmarks=true,bookmarksnumbered=true,
breaklinks]{hyperref}

\hypersetup{linkcolor=red, citecolor=blue, urlcolor=blue}

\usepackage[normalem]{ulem} 
\usepackage[dvipsnames]{xcolor} 
\begin{document}
\title{Production of $K^* \Sigma$ and $D^* \Sigma_c$ in pion-induced reactions off
the nucleon}
\author{Sang-Ho Kim}
\email{shkimphy@gmail.com}
\affiliation{Department of Physics and Origin of Matter and Evolution of Galaxies
(OMEG) Institute, Soongsil University, Seoul 06978, South Korea}

\date{\today}

\begin{abstract}

We investigate the mechanisms of strangeness production in the $\pi^- p \to K^{*0}
\Sigma^0$ and $\pi^- p \to K^{*+} \Sigma^-$ reactions within a hybrid Regge framework
in which effective Lagrangian vertices are combined with Reggeized exchanges.
The nonresonant background consists of $t$-channel $K$- and $K^*$-Reggeon exchanges,
$s$-channel nucleon and $\Delta$ exchanges, and $u$-channel $\Sigma$- and
$\Lambda$-Reggeon exchanges, whose roles differ markedly between the two isospin
channels.
We additionally include several $N^*$ and $\Delta^*$ resonances in the $s$ channel
and find that the $\Delta(2150)1/2^-$ resonance provides the dominant near-threshold
contribution.
The resulting total and differential cross sections and spin-density matrix elements
(SDMEs) are in good agreement with the available data.
Additional measurements near threshold ($W \lesssim$ 2.5 GeV) would be valuable for
clarifying the role of $s$-channel baryon resonances.
Within the same framework, with Regge trajectories and energy-scale parameters fixed
by a QGSM-motivated prescription, we also predict the cross sections for the
charm-production reactions $\pi^- p \to D^{*-} \Sigma_c^+$ and $\pi^- p \to D^{*0}
\Sigma_c^0$.
Their total cross sections are suppressed by approximately $4$--$5$ and $7$--$8$
orders of magnitude, respectively, compared with those of the corresponding
strangeness-production reactions.
These results may serve as useful guidance for future experiments at facilities such
as J-PARC.

\end{abstract}

\maketitle
\section{Introduction}

The study of excited nucleon states ($N^*$ resonances) provides valuable insight
into the internal structure of the nucleon and strong-interaction dynamics in the
nonperturbative regime of QCD.
Over the past decades, extensive efforts have been devoted to identifying and
understanding $N^*$ resonances using photon- and pion-induced reactions~\cite{
Ireland:2019uwn,Doring:2025sgb,Burkert:2025coj}.
Analyses of pseudoscalar-meson production channels such as $\pi N$, $\eta N$,
$\pi\pi N$, $K\Lambda$, and $K\Sigma$ have played a central role in extracting the
properties of $N^*$ resonances.
However, many predicted $N^*$ states remain missing or poorly established~\cite{
Edwards:2011jj,Thiel:2022xtb}, indicating that additional reaction channels may
provide complementary information on the baryon resonance spectrum.
In this respect, vector-meson production reactions offer a valuable opportunity to
investigate higher-mass baryon resonances.
For instance, $\rho N$~\cite{Ballam:1971yd,CLAS:2001zxv,Wu:2005wf,GlueX:2023fcq,
Wang:2025rvr} and $\omega N$~\cite{Oh:2000zi,Oh:2002rb,Titov:2002iv,Sibirtsev:2003qh,
Denisenko:2016ugz,Yu:2018ydp,Wei:2019imo} photoproduction processes have been widely
studied.

Interest in $K^*(892)$ vector-meson production has also grown in recent years.
For the $\gamma p \to K^{*+}\Lambda$ reaction, the first high-statistics measurements
of the total and differential cross sections~\cite{CLAS:2013qgi}, together with the
spin-density matrix elements (SDMEs)~\cite{CLAS:2017sgi}, were obtained by the
CLAS Collaboration at Jefferson Laboratory (JLab) and analyzed within the
Bonn–Gatchina (BnGa) partial-wave analysis framework.
These studies indicate that several $N^*$ resonances, namely $N(1875)3/2^-$,
$N(1880)1/2^+$, $N(1895)1/2^-$, $N(1900)3/2^+$, $N(2000)5/2^+$, $N(2060)5/2^-$,
$N(2100)1/2^+$, $N(2120)3/2^-$, and $N(2190)7/2^-$, contribute significantly to this
reaction, and their couplings to the $K^* \Lambda$ channel are listed in the 2018
edition of the Particle Data Group (PDG)~\cite{PDG:2018ovx}.
In addition, effective Lagrangian and Regge approaches have been employed to explore
the role of $N^*$ resonances in this reaction~\cite{Kim:2014hha,Wang:2017tpe,
Wang:2019mid,Wei:2020fmh,Tian:2025bkx}.

The $K^*\Sigma$ production channel is also of particular interest because it serves
as an isospin filter for disentangling the contributions of $N^*$ and $\Delta^*$
resonances.
For the $\gamma p \to K^{*+}\Sigma^{0},\, K^{*0}\Sigma^{+}$ reactions,
experimental data on the total and differential cross sections have been reported by
the CLAS~\cite{CLAS:2007kab,CLAS:2013qgi}, CBELSA/TAPS~\cite{CBELSATAPS:2008mpu},
and LEPS~\cite{Hwang:2012zza} Collaborations.
The interpretation of these data remains model dependent.
For example, the $s$-channel $N^*$ and $\Delta^*$ contributions were found to be
negligible in Ref.~\cite{Kim:2012pz}, whereas Refs.~\cite{Wang:2018vlv,Wang:2026ahb}
emphasized the role of the $\Delta(1905)5/2^+$ resonance.
In contrast, Refs.~\cite{Ben:2023uev,Shi:2025mvl} suggested that the $N(2080)3/2^-$
and $N(2270)3/2^-$ resonances provide dominant contributions.

Whereas $K^* \Lambda$ and $K^* \Sigma$ production have been extensively investigated
in photon-induced reactions, studies based on pion-induced reactions remain rather
limited.
Since pion-induced reactions couple strongly to baryon resonances, they offer
valuable information on the baryon resonance spectrum.
In this regard, we study the $\pi^- p \to K^* \Sigma$ reaction within a hybrid Regge
framework.
We provide results for the total and differential cross sections and SDMEs.
In particular, SDMEs encode the polarization information of the relevant hadrons and
serve as key observables for understanding the reaction mechanism.
We find that our results are consistent with the available experimental data reported
several decades ago.

\begin{figure*}[ht]
\centering
\includegraphics[scale=0.55]{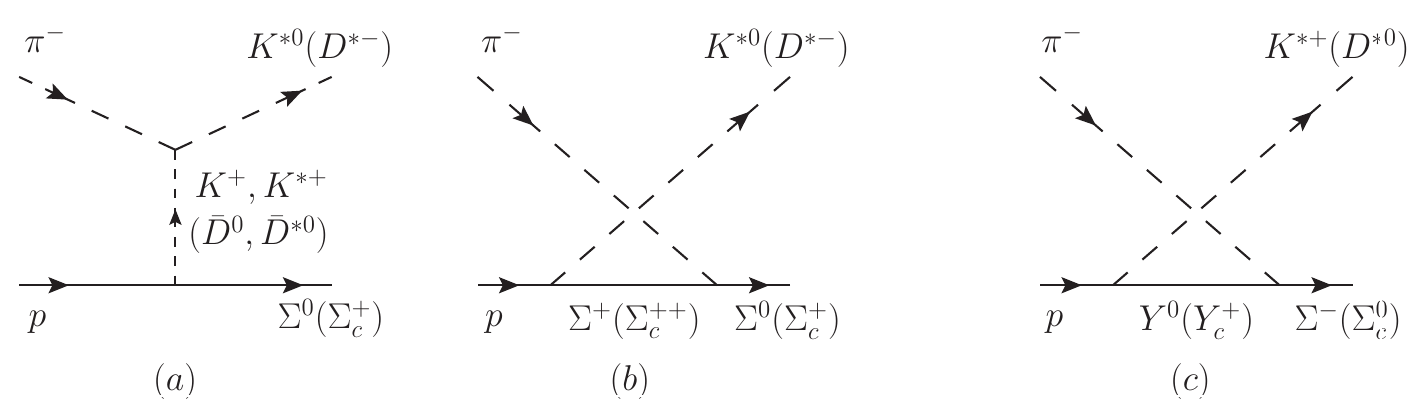}
\caption{Feynman diagrams for the $\pi^- p \to K^{*0} \Sigma^0 (D^{*-} \Sigma_c^+)$
reactions, including (a) $t$-channel pseudoscalar- and vector-Reggeon exchanges and
(b) $u$-channel $\Sigma$ ($\Sigma_c$)-Reggeon exchange.
For the $\pi^- p \to K^{*+} \Sigma^- (D^{*0} \Sigma_c^0)$ reactions, (c) denotes the
$u$-channel hyperon-Reggeon exchange, where $Y^0 = (\Lambda,\Sigma)$ and
$Y_c^+ = (\Lambda_c^+,\Sigma_c^+)$.}
\label{FIG01}
\end{figure*}
\begin{figure}[ht]
\centering
\includegraphics[scale=0.50]{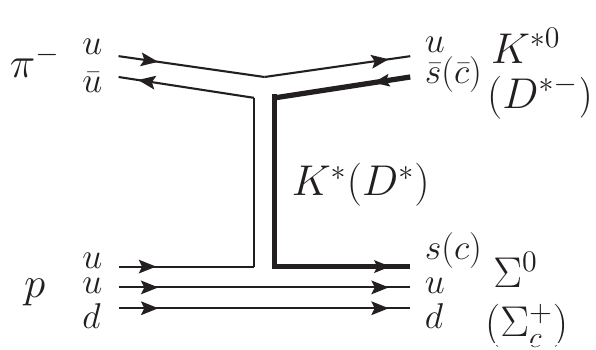}
\caption{Quark-level diagram for the $\pi^- p \to K^{*0} \Sigma^0 (D^{*-} \Sigma_c^+)$
reactions in the $t$ channel.}
\label{FIG02}
\end{figure}
A second motivation of this work is to provide predictions for the corresponding
charm-production reaction $\pi^- p \to D^* \Sigma_c$.
At J-PARC, the E50 spectrometer has been proposed to study charged open-charm hadrons
produced in $\pi^- p \to D^{*-} Y_c^{*+}$ at an incident pion momentum of 20 GeV/$c$,
where $Y_c^{*+}$ denotes excited charmed baryons, such $\Lambda_c^*$'s and
$\Sigma_c^*$'s~\cite{J-PARC:P50}.
Reliable theoretical estimates of the production cross sections are therefore
essential for evaluating the experimental feasibility~\cite{Kim:2014qha,Kim:2015ita,
Shim:2019yxn}.
In earlier studies, the author and collaborators examined the open-strangeness
($\pi^- p \to K^{(*)}\Lambda$) and open-charm ($\pi^- p \to D^{(*)}\Lambda_c^+$)
reactions~\cite{Kim:2015ita,Kim:2016imp,Kim:2017hhm} within a unified framework
motivated by the quark–gluon string model (QGSM)~\cite{Kaidalov:1982bq,
Boreskov:1983bu,Kaidalov:1986zs,Kaidalov:1994mda}.
In this framework, the interaction can be interpreted as the annihilation of a
$q\bar q$ pair followed by the formation of an intermediate color string, which
subsequently hadronizes into the observed final-state particles through planar
diagrams.
An important feature of this framework is that the Regge trajectories $\alpha(t)$
and the energy-scale parameters $s_0$ can be determined in a unified manner for both
the strangeness and charm sectors.
This formulation can therefore be applied consistently to the $\pi^- p \to K^* \Sigma$
and $\pi^- p \to D^* \Sigma_c$ reactions.

The paper is organized as follows.
Section~\ref{Sec:II} presents the hybrid Regge framework and summarizes the main
ingredients of the present model, including the coupling constants, Regge
trajectories, energy-scale parameters, and form factors.
Section~\ref{Sec:III} contains the numerical results for the total and differential
cross sections and the SDMEs.
Section~\ref{Sec:IV} gives a summary and concluding remarks.

\section{Theoretical Framework}
\label{Sec:II}

In this section, we present the hybrid Regge framework in which effective Lagrangian
vertices are combined with Reggeized exchanges.
We consider two different isospin channels:
\begin{align}
(A)\, \pi^- p \to K^{*0} \Sigma^0,
\hspace{1em}
(B)\, \pi^- p \to K^{*+} \Sigma^-,
\end{align}
for the strangeness sector, and
\begin{align}
(C)\, \pi^- p \to D^{*-} \Sigma_c^+,
\hspace{1em}
(D)\, \pi^- p \to D^{*0} \Sigma_c^0,
\end{align}
for the charm sector.
The relevant Feynman diagrams for the (A) $\pi^- p \to K^{*0} \Sigma^0$ and (C) $\pi^-
p \to D^{*-} \Sigma_c^+$ reactions are depicted in Fig.~\ref{FIG01}, including (a)
$t$-channel exchanges of Reggeized pseudoscalar- and vector-mesons and (b) $u$-channel
exchanges of Reggeized hyperons.
For the (B) $\pi^- p \to K^{*+} \Sigma^-$ and (D) $\pi^- p \to D^{*0} \Sigma_c^0$
reactions, the $t$-channel contribution is absent because no mesonic state carries
the required flavor quantum numbers.
The corresponding (c) $u$-channel exchanges of Reggeized hyperons are shown in
Fig~\ref{FIG01}.

The quark-level diagram corresponding to Fig.~\ref{FIG01}(a) is shown in
Fig.~\ref{FIG02}.
The open-charm reaction $\pi^- p \to D^{*-} \Sigma_c^+$ is constructed from the
open-strangeness reaction $\pi^- p \to K^{*0} \Sigma^0$ by replacing the strange quark
in the exchanged line with a charm quark ($s \to c$).
In this planar-diagram picture, we adopt a QGSM-motivated prescription for fixing the
Regge parameters.

\subsection{Strangeness production: $\pi^- p \to K^* \Sigma$}
\label{Sec:II-1}

We begin with the $t$-channel contribution to the reaction shown in
Fig.~\ref{FIG01}(a):
\begin{align}
\pi^- (k_1) + p (p_1) \to K^*(k_2) + \Sigma (p_2).
\end{align}
The effective Lagrangians for the meson-meson interaction vertices are given by
\begin{align}
\mathcal L_{K^* K \pi} = &\, -ig_{K^* K \pi}
( \bar K \partial^\mu \bm{\tau}\cdot\bm{\pi} K^*_\mu -
\bar K_\mu^* \partial^\mu \bm{\tau}\cdot\bm{\pi} K ),
\cr
\mathcal L_{K^* K^* \pi} = &\, {g_{K^* K^* \pi}}
\varepsilon^{\mu\nu\alpha\beta}
\partial_\mu \bar K_\nu^* \bm{\tau}\cdot\bm{\pi} \partial_\alpha K_\beta^*,
\label{eq:Lag1}
\end{align}
where $K^*$, $K$, and $\pi$ stand for the fields of $K^*(892,1^-)$, $K(494,0^-)$,
and $\pi(139,0^-)$ mesons, respectively.
The coupling constants are given by
\begin{align}
g_{K^* K \pi} = 6.56,\,\,\,  g_{K^* K^* \pi} = 7.45~\mathrm{GeV}^{-1},
\end{align}
where $g_{K^* K \pi}$ is determined from the branching ratio $\mathrm{Br} (K^* \to K
\pi) \simeq 100\,\%$ and total width $\Gamma_{K^*}$ = 51.4 MeV~\cite{PDG:2024cfk},
using
\begin{align}
\Gamma (K^* \to K \pi) = \frac{p_K^3}{8 \pi M_{K^*}^2} g_{K^* K \pi}^2,
\label{eq:DWidth-1}
\end{align}
where $p_K$ is the magnitude of the kaon three-momentum in the $K^*$ rest frame.
$M_h$ is the mass of hadron $h$.
The coupling $g_{K^* K^* \pi}$ is estimated from the anomalous VVP interaction, using
SU(3) flavor symmetry and vector-meson dominance.

The effective Lagrangians for the meson-baryon interaction vertices are written as
\begin{align}
\mathcal L_{K N \Sigma} &= -i
g_{K N \Sigma} \bar N \gamma_5 \bm{\tau} \cdot \bm{\Sigma} K + \mathrm{H.c.},
\cr
\mathcal L_{K^* N \Sigma} &= 
-g_{K^* N \Sigma} \bar N \biggl[ \gamma_\mu \bm{\tau} \cdot \bm{\Sigma} -
\cr &\,
\frac{\kappa_{K^* N \Sigma}}{M_N+M_\Sigma} \sigma_{\mu\nu} \bm{\tau} \cdot \bm{\Sigma}
\partial^\nu \biggr]
K^{*\mu} + \mathrm{H.c.},
\label{eq:Lag2}
\end{align}
where $N$ and $\Sigma$ denote the nucleon and $\Sigma(1189,1/2^+)$ baryon fields,
respectively.
We adopt the averaged values from the Nijmegen potential
model~\cite{Rijken:1998yy,Stoks:1999bz}:
\begin{align}
g_{K N \Sigma} = &\, 4.70,
\cr
g_{K^* N \Sigma} = &\, -2.99, \,\,\, \kappa_{K^* N \Sigma} = -0.910.
\label{eq:Coupl1}
\end{align}

Within the hybrid Regge framework, the $t$-channel amplitudes are obtained by
replacing the Feynman propagators of the exchanged $K$ and $K^*$ mesons with the
Regge propagators corresponding to their Regge trajectories~\cite{Kim:2015ita,
Titov:2008yf},
\begin{align}
T_K  (s,t) = &\, \mathcal M_K(s,t)
\left( \frac{s}{s_K^{\pi N : K^* \Sigma}} \right)^{\alpha_K(t)}
\cr
& \times \Gamma (-\alpha_K(t)) \alpha_K' F_{PS}(t),
\cr
T_{K^*} (s,t) = &\, \mathcal M_{K^*}(s,t)
\left( \frac{s}{s_{K^*}^{\pi N : K^* \Sigma}} \right)^{\alpha_{K^*}(t)-1}
\cr
& \times \Gamma (1-\alpha_{K^*}(t)) \alpha_{K^*}' F_V(t),
\label{eq:RegAmpl1}
\end{align}
where the amplitudes $\mathcal M_K$ and $\mathcal M_{K^*}$ are derived from the
Lagrangians given in Eq.~(\ref{eq:Lag1}), respectively,
\begin{align}
\mathcal M_{K}^\mu = &\,
I_K i g_{K^* K \pi} g_{K N \Sigma} \gamma_5 k_1^\mu,
\cr
\mathcal M_{K^*}^\mu = &\,
I_{K^*} g_{K^* K^* \pi} g_{K^* N \Sigma}
\epsilon^{\mu\nu\alpha\beta}
\cr & \times
\left [ \gamma_\nu - \frac{i\kappa_{K^* N \Sigma}}{M_N + M_\Sigma}
\sigma_{\nu \lambda}(k_2-k_1)^\lambda \right ] k_{2 \alpha} k_{1 \beta},
\label{eq:Ampl1}
\end{align}
where the full amplitude is given by
\begin{align}
\mathcal M = \varepsilon_\mu^* \bar{u}_\Sigma \, \mathcal M^\mu \,u_N.
\end{align}
Here $u_N$ and $u_\Sigma$ are the Dirac spinors of the initial nucleon and final
$\Sigma$ baryon, respectively, normalized as $\bar u_B u_B =1$.
The vector $\varepsilon_\mu$ is the polarization of the outgoing $K^*$ meson and
$I_{K(K^*)}$ = $\sqrt{2}$ is the isospin factor.

The form factor in Eq.~(\ref{eq:RegAmpl1}) accounts for the finite spatial extent of
hadrons
\begin{align}
F_{PS(V)}(t) = \left( \frac{\Lambda_{PS(V)}^2}{\Lambda_{PS(V)}^2 - t} \right)^2.
\label{eq:FormFac_t}
\end{align}
The cutoff masses are fixed by fitting the available data: $\Lambda_{PS}$ = 0.85
GeV and $\Lambda_V$ = 1.35 GeV.

The differential cross section is expressed as
\begin{align}
\frac{d\sigma}{dt} = \frac{M_N M_\Sigma}{16 \pi (p_{\mathrm{c.m.}})^2 s}
\frac{1}{2}\sum_{\lambda_V,s_f,s_i}|T|^2,
\label{eq:Def:dsdt}
\end{align}
where $p_{\mathrm{c.m.}}$ is the $\pi$ beam momentum in the center-of-mass (c.m.)
frame.
The indices $s_i$, $s_f$, and $\lambda_V$ denote the helicities of the initial
nucleon, final $\Sigma$ baryon, and $K^*$ meson, respectively.
At high energy and small momentum transfer, the cross section exhibits the Regge
behavior:
\begin{align}
\frac{d\sigma}{dt}(s \to \infty, t \to 0) \propto s^{2\alpha(0)-2}.
\label{eq:Asym:dsdt}
\end{align}

We follow Ref.~\cite{Brisudova:1999ut} to determine the $K$- and $K^*$-Regge
trajectories appearing in Eq.~(\ref{eq:RegAmpl1}).
In that approach, the so-called ``square-root'' trajectory is employed,
\begin{align}
\alpha(t) = \alpha(0) + \gamma [ \sqrt{T} - \sqrt{T-t} ],
\label{eq:SRTraj}
\end{align}
where $\gamma$ denotes a universal slope parameter and $T$ is a trajectory-dependent
scale.
For small momentum transfer, $-t \ll T$, this expression reduces to an approximately
linear form,
\begin{align}
\alpha(t) = \alpha(0) + \alpha' t,
\label{eq:LinTraj}
\end{align}
with the slope parameter $\alpha'=\gamma/(2\sqrt{T})$.
The values of $\sqrt{T}$ associated with the $\pi$- and $\rho$-Regge trajectories
were extracted in Ref.~\cite{Brisudova:1999ut} as
\begin{align}
\sqrt{T_\pi}  &= 2.82 \pm 0.05 \, \mathrm{GeV},
\cr
\sqrt{T_\rho} &= 2.46 \pm 0.03 \, \mathrm{GeV},
\label{UnivPara}
\end{align}
with $\gamma = 3.65 \pm 0.05 \, \mathrm{GeV}^{-1}$ obtained from the $\rho$
spectrum.
Using the same procedure, we determine the corresponding $\sqrt{T}$ values for the
$K$- and $K^*$-Regge trajectories.

The $\eta_s$- and $\phi$-Regge trajectories are then obtained by applying the
additivity rules for the intercepts and inverse slopes,
\begin{align}
2\alpha_{\bar s q}(0) = &\, \alpha_{\bar{q}q}(0) + \alpha_{\bar{s}s}(0),
\cr
2/\alpha'_{\bar s q} = &\, 1/\alpha'_{\bar q q} + 1/\alpha'_{\bar s s},
\label{eq:TrajRela1}
\end{align}
where the $\alpha_{\bar q q}(t)$, $\alpha_{\bar s q}(t)$, and $\alpha_{\bar s s}(t)$
denote the trajectories of the $\pi$, $K$, and $\eta_s$ in the pseudoscalar sector,
and of the $\rho$, $K^*$, and $\phi$ in the vector sector, respectively.
All resulting Regge-trajectory parameters are summarized in Table~\ref{TAB:1}.
\begin{table}[h]
\begin{tabular}{cccc}
\hline\hline
&$\alpha(0)$
&\hspace{1.1em}$\sqrt{T}\,[\mathrm{GeV}]$\hspace{1.1em}
&$\alpha'\,[\mathrm{GeV}^{-2}]$
\\\hline
$\bar{q}q(\pi)$     \hspace{5.0em}& -0.0118 & 2.82 & 0.647   \\
$\bar{s}q(K)$       \hspace{5.0em}& -0.151  & 2.96 & 0.617   \\
$\bar{s}s(\eta_s)$  \hspace{5.0em}& -0.291  & 3.10 & 0.589   \\
\hline
$\bar{q}q(\rho)$ \hspace{5.0em}& 0.55  & 2.46 & 0.742   \\
$\bar{s}q(K^*)$  \hspace{5.0em}& 0.414 & 2.58 & 0.707   \\
$\bar{s}s(\phi)$ \hspace{5.0em}& 0.27  & 2.70 & 0.676   \\
\hline\hline
\end{tabular}
\caption{Pseudoscalar- and vector-meson trajectories in the strangeness
sector~\cite{Brisudova:1999ut,Titov:2008yf}.}
\label{TAB:1}
\end{table}

Once the Regge trajectories are fixed, the energy-scale parameters
$s_K^{\pi N : K^* \Sigma}$ and $s_{K^*}^{\pi N : K^* \Sigma}$ in Eq.~(\ref{eq:RegAmpl1}) are
determined from the corresponding scale parameters of the diagonal transitions
$\pi N \to \pi N (s^{\pi N})$ and $K^* \Sigma \to K^* \Sigma (s^{K^* \Sigma})$~\cite{
Boreskov:1983bu,Kaidalov:1986zs,Kaidalov:1994mda},
\begin{align}
(s_K^{\pi N : K^* \Sigma})^{2(\alpha_K(0))}
= &\, (s^{\pi N})^{\alpha_\pi(0)} \times (s^{K^* \Sigma})^{\alpha_{\eta_s}(0)},
\cr
(s_{K^*}^{\pi N : K^* \Sigma})^{2(\alpha_{K^*}(0)-1)}
= &\, (s^{\pi N})^{\alpha_\rho(0)-1} \times (s^{K^* \Sigma})^{\alpha_\phi(0)-1},
\label{eq:EneScaPara1}
\end{align}
where $s^{ab}$ is proportional to the product of total transverse masses of the
constituent quarks in hadrons $a$ and $b$
\begin{align}
s^{ab} = \left( \sum_i m_{\perp i} \right)_a \left( \sum_j m_{\perp j} \right)_b.
\end{align}
We adopt $m_{\perp q} \simeq 0.5$ GeV, $m_{\perp s} \simeq 0.6$ GeV, and $m_{\perp c}
\simeq 1.6$ GeV, which lead to
\begin{align}
& s^{\pi N} = 1.5, \,\,\, s^{K^* \Sigma} = 1.76,
\cr
& s_K^{\pi N : K^* \Sigma} = 1.75, \,\,\,
s_{K^*}^{\pi N : K^* \Sigma} = 1.66,
\end{align}
in units of $\mathrm{GeV}^2$.
We point out that, in our model, the Regge parameters are taken to be identical for
the $K^* \Sigma$ and $K^* \Lambda$ production channels.
Accordingly, we set $s^{K^* \Sigma} = s^{K^* \Lambda}$ and $s_{K (K^*)}^{\pi N : K^* \Sigma} =
s_{K (K^*)}^{\pi N : K^* \Lambda}$.
The same assumption is applied to the charm-production case discussed below.

\subsection{Charm production: $\pi^- p \to  D^* \Sigma_c$}
\label{Sec:II-2}

We extend the strangeness-sector analysis of $\pi^- p \to K^* \Sigma$ to the
charm-production process by replacing strange hadrons with their charmed
counterparts.
The $t$-channel Regge amplitudes for the $\pi^- p \to D^* \Sigma_c$ reaction are
obtained by replacing $K \to D(1869,0^-)$ and $K^* \to D^*(2010,1^-)$ in the exchanged
propagators, and $K^* \to D^*(2010,1^-)$ and $\Sigma \to \Sigma_c (2455,1/2^+)$
in the final states [Fig~\ref{FIG01}(a)]~\cite{Kim:2015ita}:
\begin{align}
T_D (s,t) = &\, \mathcal M_D(s,t)
\left( \frac{s}{s_D^{\pi N : D^* \Sigma_c}} \right)^{\alpha_D(t)}
\cr
& \times \Gamma (-\alpha_D(t)) \alpha_D' F_{PS}(t),
\cr
T_{D^*} (s,t) = &\, \mathcal M_{D^*}(s,t)
\left( \frac{s}{s_{D^*}^{\pi N : D^* \Sigma_c}} \right)^{\alpha_{D^*}(t)-1}
\cr
& \times \Gamma (1-\alpha_{D^*}(t)) \alpha_{D^*}' F_V(t).
\label{eq:RegAmpl2}
\end{align}
The $D$- and $D^*$-Regge trajectories are determined in the same manner as in the
strangeness sector.
The corresponding trajectory parameters are listed in Table~\ref{TAB:2}~\cite{
Brisudova:1999ut}.
\begin{table}[h]
\begin{tabular}{cccc}
\hline\hline
&$\alpha(0)$
&\hspace{1.1em}$\sqrt{T}\,[\mathrm{GeV}]$\hspace{1.1em}
&$\alpha'\,[\mathrm{GeV}^{-2}]$
\\\hline
$\bar{q}q(\pi)$    \hspace{6.0em}& -0.0118  & 2.82 & 0.647   \\
$\bar{c}q(D)$      \hspace{6.0em}& -1.61105 & 4.16 & 0.439   \\
$\bar{c}c(\eta_c)$ \hspace{6.0em}& -3.2103  & 5.49 & 0.332   \\
\hline
$\bar{q}q(\rho)$   \hspace{6.0em}&  0.55 & 2.46 & 0.742   \\
$\bar{c}q(D^*)$    \hspace{6.0em}& -1.02 & 3.91 & 0.467   \\
$\bar{c}c(J/\psi)$ \hspace{6.0em}& -2.60 &5.36  & 0.340   \\
\hline\hline
\end{tabular}
\caption{Pseudoscalar- and vector-meson trajectories in the charm
sector~\cite{Brisudova:1999ut}.}
\label{TAB:2}
\end{table}

Equation~(\ref{eq:EneScaPara1}) is generalized to the charm sector as follows~\cite{
Boreskov:1983bu,Kaidalov:1986zs,Kaidalov:1994mda}:
\begin{align}
(s_D^{\pi N : D^* \Sigma_c})^{2 \alpha_D(0)}
= &\, (s^{\pi N})^{\alpha_\pi(0)} \times (s^{D^* \Sigma_c})^{\alpha_{\eta_c} (0)},
\cr
(s_{D^*}^{\pi N : D^* \Sigma_c})^{2(\alpha_{D^*}(0)-1)}
= &\, (s^{\pi N})^{\alpha_\rho(0)-1} \times (s^{D^* \Sigma_c})^{\alpha_{J/\psi}(0)-1},
\label{eq:EneScaPara2}
\end{align}
where
\begin{align}
& s^{\pi N} = 1.5, \,\,\, s^{D^* \Sigma_c} = 5.46,
\cr
& s_D^{\pi N : D^* \Sigma_c} = 5.43, \,\,\,
s_{D^*}^{\pi N : D^* \Sigma_c} = 4.75,
\end{align}
in units of $\mathrm{GeV}^2$.

The meson-meson interaction Lagrangians corresponding to Eq.~(\ref{eq:Lag1}) are
then written as
\begin{align}
\mathcal L_{D^* D \pi} = &\, -ig_{D^* D \pi}
( \bar D \partial^\mu \pi D^*_\mu - \bar D_\mu^* \partial^\mu \pi D ),
\cr
\mathcal L_{D^* D^* \pi} = &\, {g_{D^* D^* \pi}}
\varepsilon^{\mu\nu\alpha\beta}
\partial_\mu \bar D_\nu^* \,\pi\, \partial_\alpha D_\beta^*.
\label{eq:Lag3}
\end{align}
The coupling constant $g_{D^* D \pi} $ is determined from the branching ratios~\cite{
PDG:2024cfk}
\begin{align}
\mathrm{Br} (D^{*\pm} \to D^0 \pi^\pm) &= 67.7 \pm 0.5\,\%,
\cr
\mathrm{Br} (D^{*\pm} \to D^\pm \pi^0) &= 30.7 \pm 0.5\,\%,
\label{eq:BR}
\end{align}
and total width $\Gamma_{D^{*\pm}}$ = 83.4 keV~\cite{PDG:2024cfk}, using
\begin{align}
\Gamma (D^{*\pm} \to D \pi) = \frac{g_{D^* D \pi}^2}{24 \pi} \frac{p_D^3}{M_{D^*}^2}.
\label{eq:DWidth-2}
\end{align}
Here $p_D$ is the magnitude of the $D$ three-momentum in the $D^*$ rest frame.
This yields
\begin{align}
g_{D^{*\pm} D^0 \pi^\mp} = 16.8, \,\,\, g_{D^{*\pm} D^\mp \pi^0} = 11.7.
\label{eq:Coupl2}
\end{align}
In our numerical calculations, we use $g_{D^* D \pi}$ = $g_{D^{*\pm} D^0 \pi^\mp}$ and
set $I_{D(D^*)}$ = 1 in Eq.~(\ref{eq:Ampl1}).
Note that the isospin structure $\bm{\tau}\cdot\bm{\pi}$ appearing in
Eq.~(\ref{eq:Lag1}) is not involved here.
The strength of the $D^* D^* \pi$ vertex is estimated using heavy-quark spin symmetry,
which relates it to $g_{D^* D \pi}$ through $g_{D^* D^* \pi} = g_{D^* D \pi} /
\sqrt{M_{D} M_{D^*}}$~\cite{Colangelo:2003sa,Guo:2010ak}:
\begin{align}
g_{D^{*\pm} D^{*0} \pi^\mp} &= 8.68\, \mathrm{GeV}^{-1},
\cr
g_{D^{*\pm} D^{*\mp} \pi^0} &= 6.03\, \mathrm{GeV}^{-1},
\label{eq:Coupl3}
\end{align}
and we use $g_{{D^*}D^* \pi} = g_{D^{*\pm} D^{*0} \pi^\mp}$.

For the meson-baryon couplings given in Eq.~(\ref{eq:Lag2}), we assume SU(4) flavor
symmetry and take them to be identical to those in the strange sector:
$g_{D N \Sigma_c} = g_{K N \Sigma}$, $g_{D^* N \Sigma_c} = g_{K^* N \Sigma}$, and
$\kappa_{D^* N \Sigma_c} = \kappa_{K^* N \Sigma}$.

\subsection{$u$-channel hyperon-Reggeon exchange}
\label{Sec:II-3}

We next consider the $u$-channel hyperon exchange shown in Figs.~\ref{FIG01}(b) and
(c).
In analogy with the $t$-channel treatment, the hyperon exchange is Reggeized to
describe the high-energy behavior of the amplitude.

The effective Lagrangians for the $\pi Y \Sigma$ interactions are written as
\begin{align}
\mathcal{L}_{\pi \Lambda \Sigma} &= - i g_{\pi \Lambda \Sigma} \bar \Lambda
\gamma_5 \bm{\Sigma} \cdot \bm{\pi} + \mathrm{H.c.},
\cr
\mathcal{L}_{\pi \Sigma \Sigma} &= - i g_{\pi \Sigma \Sigma} \bar \Sigma \gamma_5
\bm{T} \cdot \bm{\pi} \Sigma,
\label{eq:Lag4}
\end{align}
where $\bm{T}$ denotes the $3 \times 3$ isospin-1 generator.
The coupling constants are taken from the Nijmegen potential model~\cite{
Rijken:1998yy,Stoks:1999bz}:
\begin{align}
g_{\pi \Lambda \Sigma} = 11.9, \,\,\, g_{\pi \Sigma \Sigma} = 11.7. 
\label{eq:Coupl4}
\end{align}

The Reggeized $u$-channel amplitude for $\pi^- p \to K^* \Sigma$ is constructed
as~\cite{Kim:2015ita}
\begin{align}
T_Y (s,u)  &= \mathcal M_Y(s,u)
\left( \frac{s}{s_Y^{\pi N : K^* \Sigma}} \right)^{\alpha_Y(u)-\frac{1}{2}} 
\cr   & \times
\Gamma \left( \frac{1}{2}-\alpha_Y(u) \right) \alpha_Y' F_Y(u), 
\label{eq:RegAmpl3}
\end{align}
with the amplitude
\begin{align}
\mathcal M_Y^\mu &=
I_Y i g_{K^* N Y} g_{\pi Y \Sigma}
\gamma_5 (\rlap{/}{p_2}-\rlap{/}{k_1}+M_Y)
\cr & \times
\left [ \gamma^\mu - \frac{i\kappa_{K^* N Y}}{M_N + M_Y} \sigma^{\mu\nu} k_{2 \nu}
\right ],
\label{eq:Ampl3}
\end{align}
and $\mathcal M = \varepsilon_\mu^* \bar{u}_\Sigma\, \mathcal M^\mu\, u_N$.

The linear Regge trajectories of the exchanged hyperons are taken from
Ref.~\cite{Storrow:1983ct}:
\begin{align}
\alpha_\Lambda (u) &= -0.65 + 0.94 u,
\cr
\alpha_\Sigma (u) &= -0.79 + 0.87 u.
\label{eq:YTraj-1}
\end{align}
The hadronic form factors in the $u$ channel are chosen to be phenomenological as
\begin{align}
F_\Sigma (u) &= \left( \frac{\Lambda_Y^2}{\Lambda_Y^2 - u} \right)^2,
\cr
F_\Lambda (u) &= \left( \frac{n\Lambda_\Lambda^4}{n\Lambda_\Lambda^4 + (u - M_\Lambda^2)^2}
\right)^n,
\label{eq:FormFac_u}
\end{align}
where the cutoff masses are adjusted to reproduce the backward-angle data:
$\Lambda_\Sigma$ = 0.33 GeV and $\Lambda_\Lambda$ = 0.53 GeV with $n$ = 1.
The Reggeized $u$-channel amplitude satisfies the asymptotic behavior 
\begin{align}
\frac{d\sigma}{du}(s \to \infty, u \to 0) \propto s^{2\alpha(0)-2}.
\label{eq:Asym:dsdu}
\end{align}

The $u$-channel Regge amplitude for the $\pi^- p \to D^* \Sigma_c$ reaction is
obtained by replacing the strange hadrons with their charmed counterparts, following
the same procedure as outlined in Sec.~\ref{Sec:II-2}.
Thus we have~\cite{Kim:2015ita}
\begin{align}
T_{Y_c} (s,u)  &= \mathcal M_{Y_c}(s,u)
\left( \frac{s}{s_{Y_c}^{\pi N : D^* \Sigma_c}} \right)^{\alpha_{Y_c}(u)-\frac{1}{2}} 
\cr   & \times
\Gamma \left( \frac{1}{2}-\alpha_{Y_c}(u) \right) \alpha_{Y_c}' F_{Y_c}(u).
\label{eq:RegAmpl4}
\end{align}

The corresponding coupling constants are
\begin{align}
g_{\pi \Lambda_c \Sigma_c} = 20.4, \,\,\, g_{\pi \Sigma_c \Sigma_c} = 8.0. 
\label{eq:Coupl4}
\end{align}
The coupling $g_{\pi \Lambda_c \Sigma_c}$ is determined from the branching ratio
$\mathrm{Br} (\Sigma_c \to \Lambda_c \pi) \simeq$ 100\,\% and $\Gamma_{\Sigma_c}
\simeq$ 2 MeV~\cite{PDG:2024cfk},
\begin{align}
\Gamma (\Sigma_c \to \Lambda_c \pi) =
\frac{g_{\pi \Lambda_c \Sigma_c}^2}{4 \pi}
\frac{p_{\Lambda_c} (E_{\Lambda_c}-M_{\Lambda_c})}{M_{\Sigma_c}},
\label{eq:DWidth-4}
\end{align}
where $p_{\Lambda_c}$ is the magnitude of the $\Lambda_c$ three-momentum in the
$\Sigma_c$ rest frame and $E_{\Lambda_c} = \sqrt{M_{\Lambda_c}^2 + p_{\Lambda_c}^2}$.
The coupling $g_{\pi \Sigma_c \Sigma_c}$ is taken from light cone QCD sum rules~\cite{
Azizi:2010sy}.

The Regge trajectories of the charmed hyperons are given by~\cite{Titov:2008yf}
\begin{align}
\alpha_{\Lambda_c} (u) &= -2.09 + 0.56 u,
\cr
\alpha_{\Sigma_c} (u) &= -2.23 + 0.53 u.
\label{eq:YTraj-2}
\end{align}

Since the QGSM applies only to the planar diagram~\cite{Kaidalov:1982bq,
Boreskov:1983bu,Kaidalov:1986zs, Kaidalov:1994mda}, it does not uniquely determine
the energy scale parameters $s_Y^{\pi N : K^* \Sigma}$ and $s_{Y_c}^{\pi N : D^* \Sigma_c}$ in
the $u$ channel.
Instead, we employ a phenomenological relation between the energy-scale parameter
$s_0$ and the reaction threshold $s_{\mathrm{th}}$, which has been found to hold
consistently in both the strangeness and charm sectors.
Using the $t$-channel values as reference, we define $s_0 / \sqrt{s_{\mathrm{th}}} =
\beta$.
This gives $s_{K(K^*)}^{\pi N : K^* \Sigma} / \sqrt{s_{\mathrm{th}}^s} \simeq$ 1.2 GeV and
$s_{D(D^*)}^{\pi N : D^* \Sigma_c} / \sqrt{s_{\mathrm{th}}^c} \simeq$ 0.9 GeV, with
$\sqrt{s_{\mathrm{th}}^s} = M_{K^*} + M_\Sigma$ = 2.08 GeV and $\sqrt{s_{\mathrm{th}}^c} =
M_{D^*} + M_{\Sigma_c}$ = 4.46 GeV.
Interestingly, the value $\beta \simeq$ 1 GeV also appears in other reactions, such
as $\bar p p \to \bar \Sigma \Sigma, \bar \Sigma_c^+ \Sigma_c^+$ and $\bar p p \to
\bar K K, \bar D D $~\cite{Titov:2008yf}.
We therefore adopt $s_\Sigma^{\pi N : K^* \Sigma} / \sqrt{s_{\mathrm{th}}^s}$ = 1.2 GeV
and $s_{\Sigma_c}^{\pi N : D^* \Sigma_c} / \sqrt{s_{\mathrm{th}}^c}$ = 0.9 GeV for the
$u$-channel energy-scale parameters in this calculation.

\subsection{$s$-channel nucleon and $\Delta$ exchanges: $\pi^- p \to K^* \Sigma$}
\label{Sec:II-4}

\begin{figure}[ht]
\centering
\includegraphics[scale=0.55]{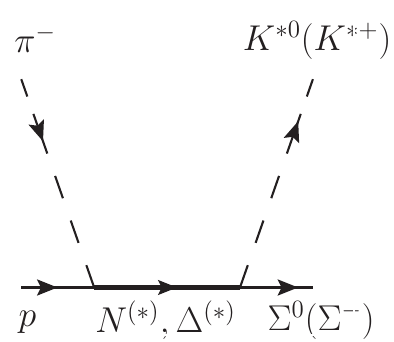}
\caption{Feynman diagram for the $\pi^- p \to K^* \Sigma$ reaction via $s$-channel
$N^{(*)}$ and $\Delta^{(*)}$ exchanges.}
\label{FIG03}
\end{figure}

Besides the dominant $t$- and $u$-channel contributions, the $s$-channel nucleon
and $\Delta$ exchanges shown in Fig.~\ref{FIG03} may also play a role in the
$\pi^- p \to K^* \Sigma$ reaction and are therefore included for completeness.

The relevant effective Lagrangians are written as
\begin{align}
\mathcal{L}_{\pi N N} &= - i g_{\pi N N} \bar N \gamma_5 \bm{\tau}\cdot\bm{\pi} N,
\cr
\mathcal{L}_{\pi N \Delta} &= \frac{g_{\pi N \Delta}}{M_\pi} \bar \Delta^\mu
\mathit{O}_{\mu\nu}(Z) \bm{I} \left( \frac32, \frac12 \right) \cdot \partial^\nu
\bm{\pi} N + \mathrm{H.c.},
\cr
\mathcal{L}_{K^* \Sigma \Delta} &= -i\frac{g_{K^* \Sigma \Delta}}{2M_{K^*}} \bar \Delta^\mu
\gamma^\nu \gamma_5 \bm{I} \left( \frac32, \frac12 \right) \cdot \bm{\Sigma}
K^*_{\mu\nu} + \mathrm{H.c.},
\label{eq:Lag4}
\end{align}
where $\Delta$ denotes the $\Delta (1232,3/2^+)$ baryon field and the isospin
transition matrix reads
\begin{align}
\bm{I} \left( \frac32, \frac12 \right) \cdot \bm{\pi} =
- I_{3/2,1/2}^{(+1)} \pi^+ + I_{3/2,1/2}^{(0)} \pi^0 + I_{3/2,1/2}^{(-1)} \pi^-,
\end{align}
with
\begin{align}
I_{3/2,1/2}^{(+1)} &=
\frac{1}{\sqrt{6}}
\left(\begin{array}{cc}
\sqrt{6}&0\\
0&\sqrt{2}\\
0&0\\
0&0
\end{array} \right),\,\,\,
I_{3/2,1/2}^{(0)} =
\frac{1}{\sqrt{6}}
\left(\begin{array}{cc}
0&0\\
2&0\\
0&2\\
0&0
\end{array} \right), \cr
& I_{3/2,1/2}^{(-1)} =
\frac{1}{\sqrt{6}}
\left(\begin{array}{cc}
0&0\\
0&0\\
\sqrt{2}&0\\
0&\sqrt{6}
\end{array} \right).
\end{align}

The coupling constants are given by
\begin{align}
g_{\pi N N} &= 13.0,   \cr
g_{\pi N \Delta} &= 2.17,   \cr
g_{K^* \Sigma \Delta} &= -12.8,
\end{align}
where $g_{\pi N N}$ is taken from the Nijmegen potential
model~\cite{Rijken:1998yy,Stoks:1999bz}.
Taking the off-shell parameter as $\mathit{O}_{\mu\nu}(Z)=g_{\mu\nu}$, the $\Delta \to
N \pi$ decay width can be written as
\begin{align}
\Gamma (\Delta \to N \pi) =
\frac{g_{\pi N \Delta}^2}{12 \pi} \frac{p_N^3}{M_\Delta M_\pi^2} (E_N + M_N).
\label{eq:DWidth-3}
\end{align}
Then the coupling $g_{\pi N \Delta}$ is determined using $M_\Delta$ = 1232 MeV,
$\Gamma_\Delta \simeq$ 117 MeV, and $\mathrm{Br} (\Delta \to N \pi)$ =
99.4\,\%~\cite{PDG:2024cfk}.
The coupling $g_{K^* \Sigma \Delta}$ is estimated using the quark model prediction and
SU(3) flavor symmetry relation,
\begin{align}
f_{K^* \Sigma \Delta}=-\frac{2M_{K^*}}{M_\rho}f_{\rho N\Delta},
\end{align}
with $f_{\rho N\Delta}$ = 5.5.

The invariant amplitudes for the $s$-channel nucleon and $\Delta$ exchanges are then
given by
\begin{align}
\mathcal M_N^\mu &=
I_N \frac{i g_{K^* N \Sigma}}{s-M_N^2} \frac{g_{\pi N N}}{2 M_N}
\left [ \gamma^\mu - \frac{i\kappa_{K^* N \Sigma}}{M_N + M_\Sigma} \sigma^{\mu\nu}
k_{2 \nu} \right ]
\cr & \times
(\rlap{/}{q_s}+M_N) \gamma^\alpha \gamma_5 k_{1\alpha},
\cr
\mathcal M_\Delta^\mu &=
I_\Delta \frac{i}{s-M_\Delta^2+iM_\Delta\Gamma_\Delta}
\frac{g_{\pi N \Delta}}{M_\pi} \frac{g_{K^* \Delta \Sigma}}{2 M_{K^*}} \gamma_\nu \gamma_5
\cr & \times
(k_2^\alpha g^{\mu\nu} - k_2^\nu g^{\mu\alpha})                              
(\rlap{/}{q_s}+M_\Delta) \Delta_{\alpha\beta} (q_s,M_\Delta) k_1^\beta,
\label{eq:Ampl2}
\end{align}
respectively, with $\mathcal M = \varepsilon_\mu^* \bar{u}_\Sigma\, \mathcal M^\mu\, u_N$
and $q_s = k_1 + p_1$.
The spin-3/2 projection operator is given by
\begin{align}
\Delta_{\alpha\beta}(p, M) &= (\slashed{p} + M)
\biggl[ -g_{\alpha\beta} + \frac13 \gamma_\alpha \gamma_\beta
\cr
& + \frac{1}{3M} (\gamma_\alpha p_\beta - p_\alpha \gamma_\beta) + \frac{2}{3M^2} p_\alpha
p_\beta \biggr].
\label{eq:SpinProj}
\end{align}

We introduce the following form factor for the $s$-channel amplitude:
\begin{align}
F_B (s) &= \left( \frac{n \Lambda_B^4}{n \Lambda_B^4 + (s - M_B^2)^2} \right)^n,
\label{eq:FormFac_s}
\end{align}
where $B = (N,\,\Delta)$.
We take $\Lambda_{N,\,\Delta}$ = 0.65 GeV with $n$ = 1.

\subsection{$s$-channel  $N^*$ and $\Delta^*$ exchanges: $\pi^- p \to K^* \Sigma$}
\label{Sec:II-5}

\begin{table*}[ht]
\centering
\begin{tabular}{cccccccc}
\hline\hline
Resonances
&\hspace{1.2em} Rating \hspace{1.2em}
&$\Gamma_R$ [MeV]
&\hspace{1.2em} $\mathrm{Br}_{R \to \pi N}$ [\%] \hspace{1.2em}
&$|g_{\pi N R}|$
&\hspace{1.2em} $\sqrt{\Gamma_{R \to K^* \Sigma}}$ \hspace{1.2em}
&$|g_{K^* \Sigma R}|$ \\
\hline
$N(2120)\,3/2^-$ & *** & 260--360 (300) & 5--15 & 0.282
& $0.6_{-0.5}^{+1.0}$ & 0.00581 \\
$N(2190)\,7/2^-$& **** & 300--500 (400) & 10--20 & 0.0216
& $0.3_{-0.2}^{+0.4}$ & 0.0806     \\
$\Delta (2150)\,1/2^-$& * & 100--300 (300) & 6--10 & 0.582
& $4.8_{-4.2}^{+0.9} $ & 1.21 \\
$\Delta (2200)\,7/2^-$ & *** & 200--500 (350)& 2--8 & 0.0196
& $0.5_{-0.5}^{+0.9}$ & 0.369 \\
$\Delta (2350)\,5/2^-$ & * & 230--550 (350) & 4--30 & 0.0488
& $1.6_{-1.0}^{+2.4}$ & 0.289 \\
$\Delta (2390)\,7/2^+$ & * & 220--400 (300) & 3--12 & 0.00604
& $1.1_{-0.8}^{+1.3}$ & 0.992 \\
\hline\hline
\end{tabular}
\caption{Strong decays of the $N^*$ and $\Delta^*$ resonances into the
$\pi N$~\cite{PDG:2024cfk} and $K^* \Sigma$ channels~\cite{Capstick:1998uh}.
The listed couplings $g_{\pi N R}$ and $g_{K^* \Sigma R}$ are extracted from the central
values of the branching ratios $\mathrm{Br}_{R \to \pi N}$ and partial decay widths
$\sqrt{\Gamma_{R \to K^* \Sigma}}$, respectively.
$\Gamma_{R \to K^* \Sigma}$ is in unit of MeV.}
\label{TAB:3}
\end{table*}

As shown in Fig.~\ref{FIG03}, we include several $s$-channel $N^*$ and $\Delta^*$
resonances in $\pi^- p \to K^* \Sigma$, which are expected to couple strongly to both
the $\pi N$ and $K^* \Sigma$ channels.
These resonances are listed in Table~\ref{TAB:3}.

The effective Lagrangians for the $\pi N R$ interaction with resonance spins
$1/2$--$7/2$ are given by~\cite{Kim:2024mqx}
\begin{align}
\mathcal L_{\pi N R}^{(1/2^\pm)} &=
-i g_{\pi N R} \bar N \Gamma^{(\pm)} \pi R + \mathrm{H.c.},
\cr
\mathcal L_{\pi N R}^{(3/2^\pm)} &=
\frac{g_{\pi N R}}{M_\pi} \bar N \Gamma^{(\mp)} \partial_\mu \pi R^\mu + \mathrm{H.c.},
\cr
\mathcal L_{\pi N R}^{(5/2^\pm)} &=
i \frac{g_{\pi N R}}{M_\pi^2} \bar N \Gamma^{(\pm)} \partial_\mu \partial_\nu
\pi R^{\mu\nu} + \mathrm{H.c.},
\cr
\mathcal L_{\pi N R}^{(7/2^\pm)} &=
- \frac{g_{\pi N R}}{M_\pi^3} \bar N \Gamma^{(\mp)} \partial_\mu \partial_\nu
\partial_\alpha
\pi R^{\mu\nu\alpha} + \mathrm{H.c.}.
\label{eq:ResLag1}
\end{align}
The effective Lagrangians for the $K^* \Sigma R$ interaction can be written
as~\cite{Kim:2024mqx}
\begin{align}
\mathcal{L}_{K^* \Sigma R}^{(1/2^\pm)}
&= - g_{K^* \Sigma R} \bar K^{*\mu} \bar \Sigma \Gamma_\mu^{(\mp)} R + \mathrm{H.c.},
\cr
\mathcal{L}_{K^* \Sigma R}^{(3/2^\pm)}
&= - i \frac{g_{K^* \Sigma R}}{2M_N}
\bar K^{*\mu\nu} \bar \Sigma  \Gamma_\nu^{(\pm)} R_\mu + \mathrm{H.c.},
\cr
\mathcal{L}_{K^* \Sigma R}^{(5/2^\pm)}
&= \frac{g_{K^* \Sigma R}}{(2M_N)^2}
\partial^\alpha \bar K^{*\mu\nu} \bar \Sigma  \Gamma_\nu^{(\mp)} R_{\mu\alpha} +
\mathrm{H.c.},
\cr
\mathcal{L}_{K^* \Sigma R}^{(7/2^\pm)}
&= i \frac{g_{K^* \Sigma R}}{(2M_N)^3}
\partial^\alpha \partial^\beta \bar K^{*\mu\nu} \bar \Sigma  \Gamma_\nu^{(\pm)}
R_{\mu\alpha\beta} + \mathrm{H.c.},
\label{eq:ResLag2}
\end{align}
where $K^{*\mu\nu} = \partial^\mu K^{*\nu} - \partial^\nu K^{*\mu}$ and $R$ denotes the
fields of the $N^*$ and $\Delta^*$ resonances.
The projection operators corresponding to the high-spin resonances are summarized in
Ref.~\cite{Kim:2024mqx}.
Since the $N^*$ and $\Delta^*$ contributions are relevant mainly near threshold,
additional interaction terms are neglected in the present work.
The following notations are used:
\begin{align}
\Gamma^{(\pm)} = \left(
\begin{array}{c} 
\gamma_5 \\ \mathbf{1}
\end{array} \right),
\,\,\,
\Gamma_\mu^{(\pm)} = \left(
\begin{array}{c}
\gamma_\mu \gamma_5 \\ \gamma_\mu 
\end{array} \right).
\end{align}

Regarding the $R \to \pi N$ couplings, we use the information from the
PDG~\cite{PDG:2024cfk}.
The coupling constants listed in Table~\ref{TAB:3} are obtained from the central
values of the $R \to \pi N$ decay widths.
For the $R \to K^* \Sigma$ couplings, we use the quark model predictions from
Ref.~\cite{Capstick:1998uh}.
The corresponding coupling constants in Table~\ref{TAB:3} are extracted from the
central values of the $R \to K^* \Sigma$ decay widths.
After fixing the $K^* \Sigma R$ couplings from the central values listed in
Table~\ref{TAB:3}, we fit the data by varying the $\pi N R$ couplings within the
ranges allowed by the PDG decay widths.
The best fit is obtained by multiplying the central values of $g_{\pi N N^*}$ and
$g_{\pi N \Delta^*}$ by factors of 0.8 and 0.6, respectively.
For the $\Delta(2150)$ resonance, the $g_{\pi N \Delta^*}$ coupling is multiplied by a
factor of $\sqrt3$ relative to the central value.
Among the six baryon resonances considered here, only the $N(2190)7/2^-$ and
$\Delta (2150)1/2^-$ resonances require different choices for the relative signs
between $g_{\pi N R}$ and $g_{K^* \Sigma R}$.

For the $s$-channel diagrams with the higher-spin baryon resonances, we introduce the
Gaussian form factor that can suppress sufficiently the cross sections when the
energy grows:
\begin{align}
F_R (s)= \exp\left[ \frac{-(s-M_R^2)^2}{\Lambda_R^4} \right] ,
\end{align}
with $\Lambda_R = 1.0$ GeV.
Our results do not depend strongly on the cutoff mass $\Lambda_R$.

\section{Numerical Results}
\label{Sec:III}

\begin{figure*}[ht]
\centering
\includegraphics[width=7.0cm]{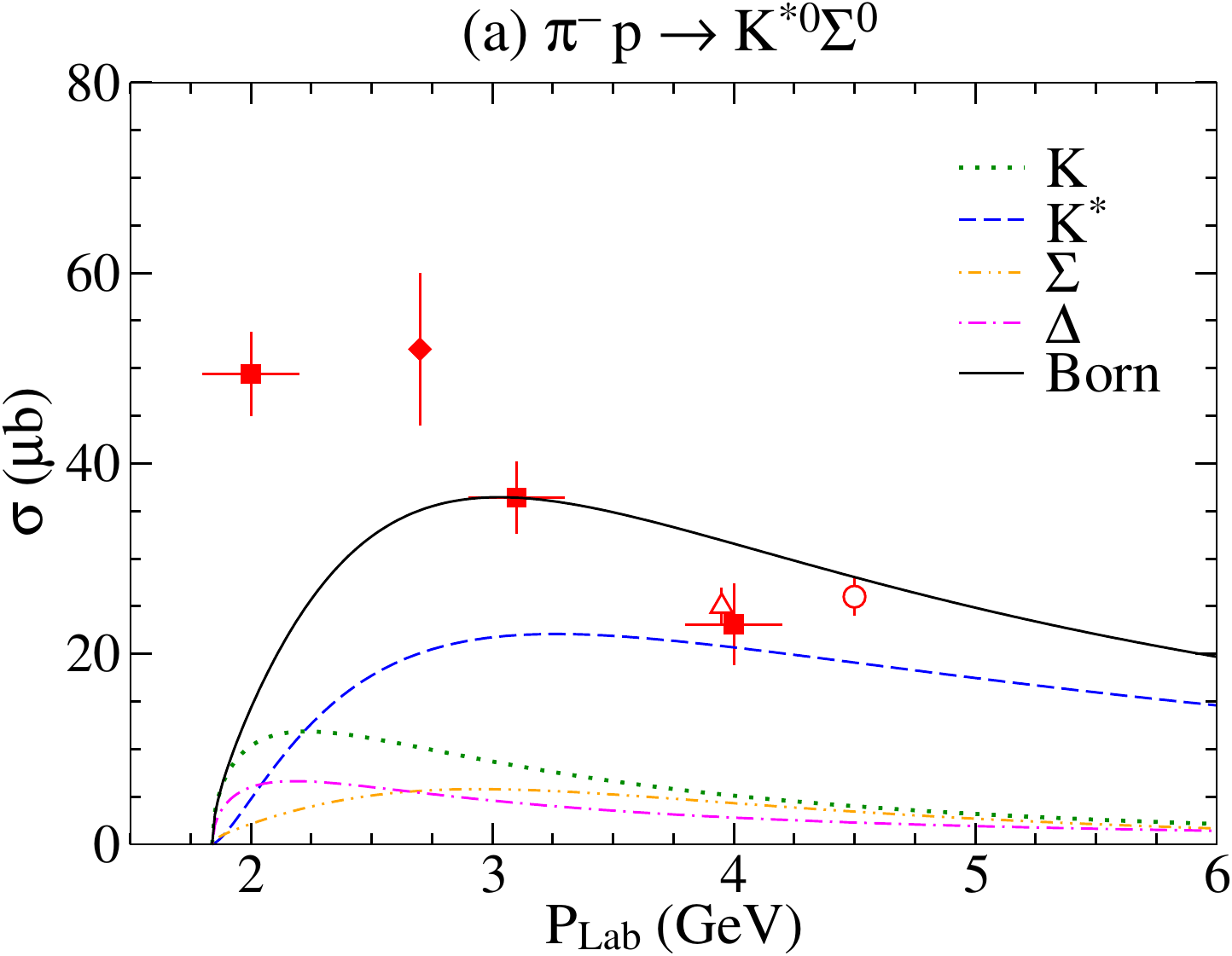} \hspace{1em}
\includegraphics[width=7.0cm]{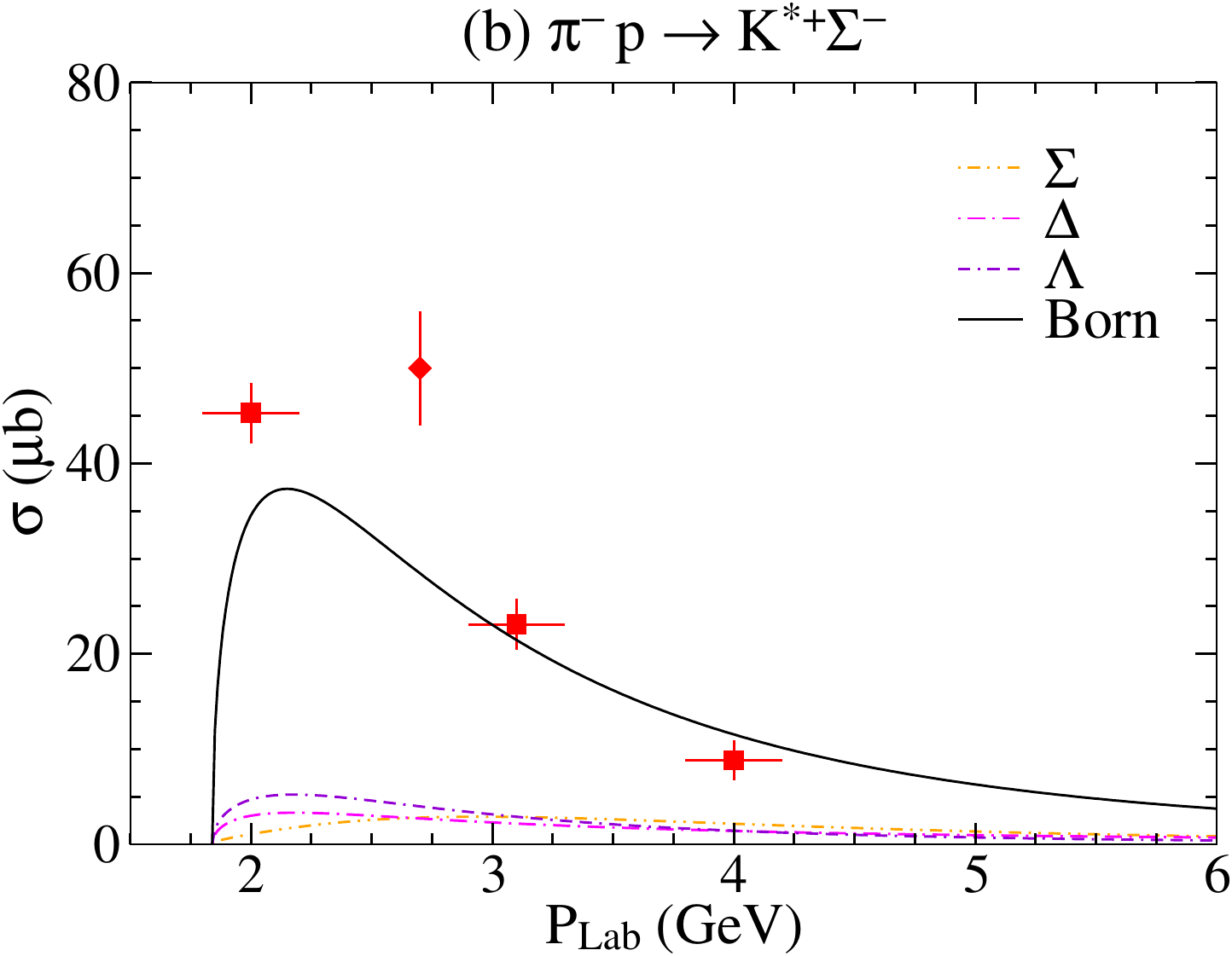}
\caption{Total cross sections as functions of $P_{\mathrm{Lab}}$ for the (a) $\pi^- p
\to K^{*0} \Sigma^0$ and (b) $\pi^- p \to K^{*+} \Sigma^-$ reactions without
resonance contributions.
The green dotted, blue dashed, orange dash-detted-detted, magenta dash-dotted, and
violet dash-dash-dotted curves represent the contributions from the $K$-, $K^*$-,
$\Sigma$-, and $\Lambda$-Reggeon exchanges, and $\Delta$ exchange, respectively.
The black solid curves denote the Born contributions.
Experimental data are taken from Refs.~\cite{Dahl:1967pg} (squares),
\cite{Miller:1965} (diamonds), \cite{CCMS:1980ysu} (open triangles), and
\cite{Crennell:1972km} (open circles).}
\label{FIG04}
\end{figure*}
\begin{figure*}[ht]
\centering
\includegraphics[width=7.0cm]{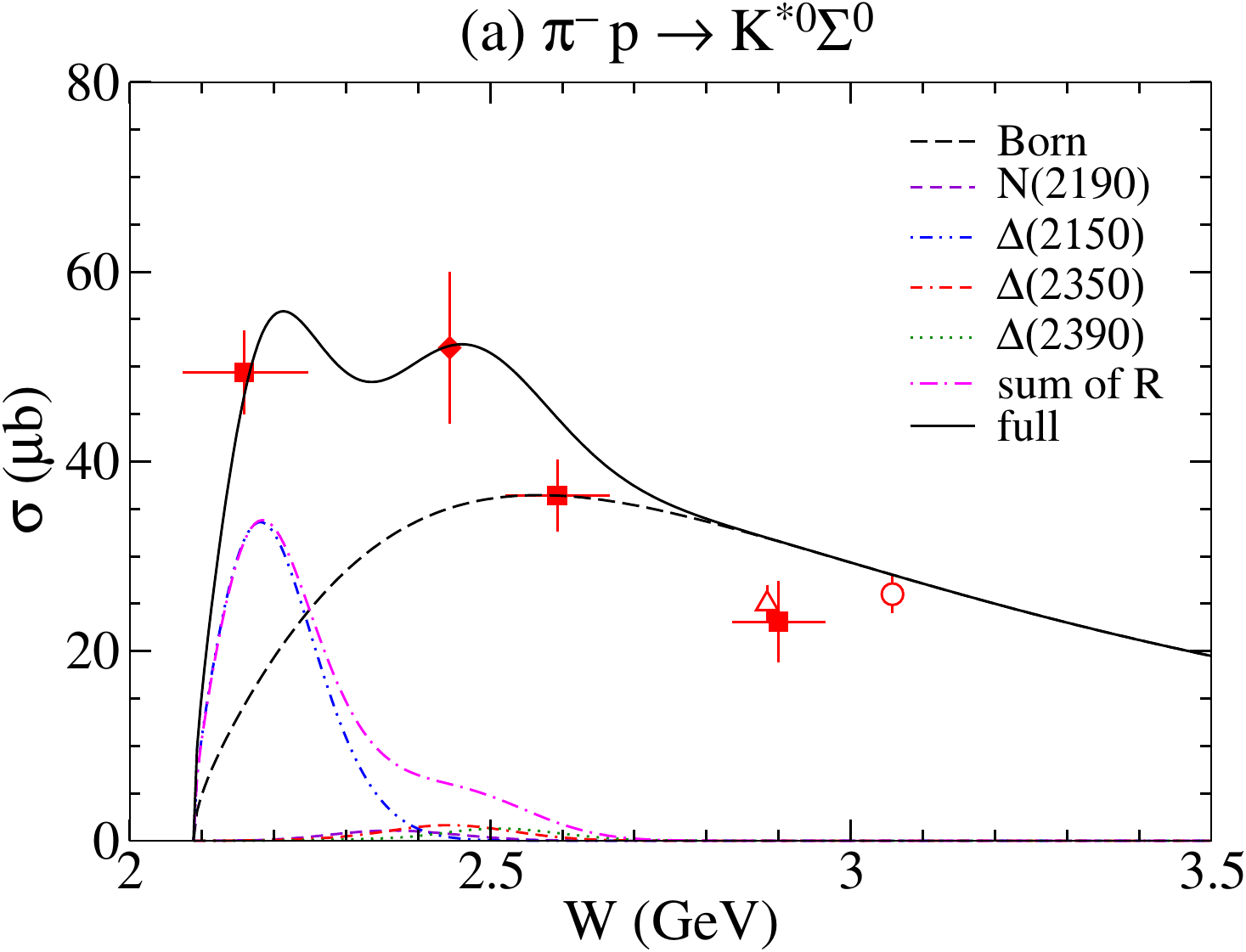} \hspace{1em}
\includegraphics[width=7.0cm]{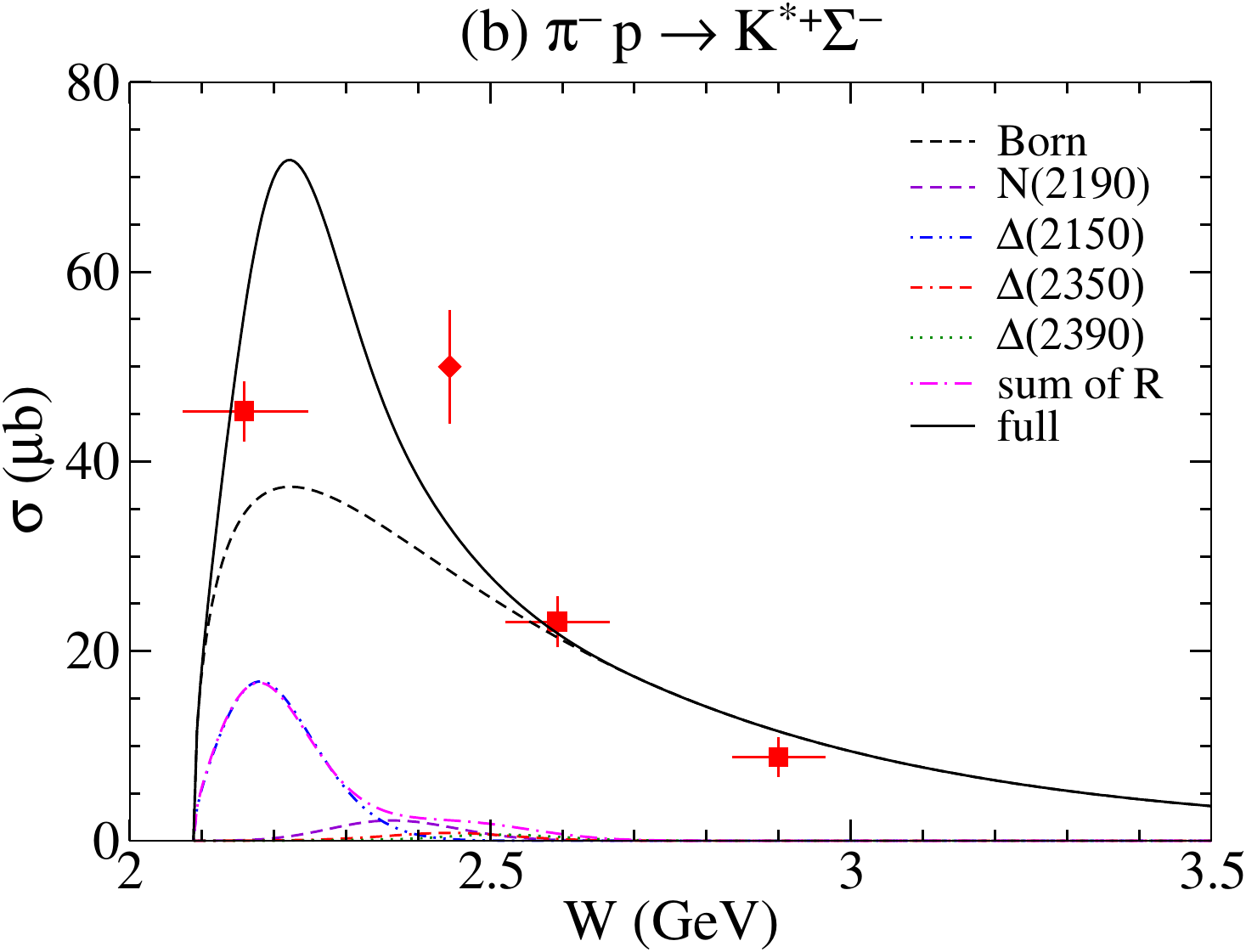}
\caption{Total cross sections as functions of $W$ for the (a) $\pi^- p \to K^{*0}
\Sigma^0$ and (b) $\pi^- p \to K^{*+} \Sigma^-$ reactions.
The black dashed and magenta dash-dotted curves represent the Born and resonance
contributions, respectively.
The black solid curves denote the full results.
Experimental data are taken from Refs.~\cite{Dahl:1967pg} (squares),
\cite{Miller:1965} (diamonds), \cite{CCMS:1980ysu} (open triangles), and
\cite{Crennell:1972km} (open circles).}
\label{FIG05}
\end{figure*}
We now present our numerical results for the strangeness- and charm-production
reactions $\pi^- p \to K^* \Sigma$ and $\pi^- p \to D^* \Sigma_c$.
We first provide the results for the strangeness production and compare with the
available experimental data~\cite{Dahl:1967pg,Miller:1965,Crennell:1972km,
CCMS:1980ysu,Abramovich:1972rq,Yaffe:1973ex,CCMS:1980mch}.
Figure~\ref{FIG04} displays the total cross sections for the two different isospin
channels, (a) $\pi^- p \to K^{*0} \Sigma^0$ and (b) $\pi^- p \to K^{*+} \Sigma^-$, as
functions of the laboratory (Lab) beam momentum $P_{\mathrm{Lab}}$.
Only the Born contributions are considered here, i.e., the nonresonant background
terms, namely the Reggeized $t$- and $u$-channel exchanges together with the
$s$-channel ground-state baryon exchanges.
For the (a) $K^{*0} \Sigma^0$ channel, the $t$-channel $K^*$-Reggeon exchange provides
the dominant contribution, particularly at high energies.
The $K$-Reggeon, $s$-channel $\Delta$, and $u$-channel $\Sigma$-Reggeon exchanges
also give non-negligible contributions.
The effect of the $s$-channel nucleon exchange remains negligible and therefore
its contribution is not shown.
For the (b) $K^{*+} \Sigma^-$ channel, the $t$-channel contribution is absent, and the
$\Delta$- and $\Sigma$-exchange terms alone are insufficient to reproduce the data.
The $u$-channel $\Lambda$-Reggeon exchange, which is allowed only in this channel,
is essential for improving the description at $P_{\mathrm{Lab}} \gtrsim$ 3 GeV.

\begin{figure*}[ht]
\centering
\includegraphics[width=\columnwidth]{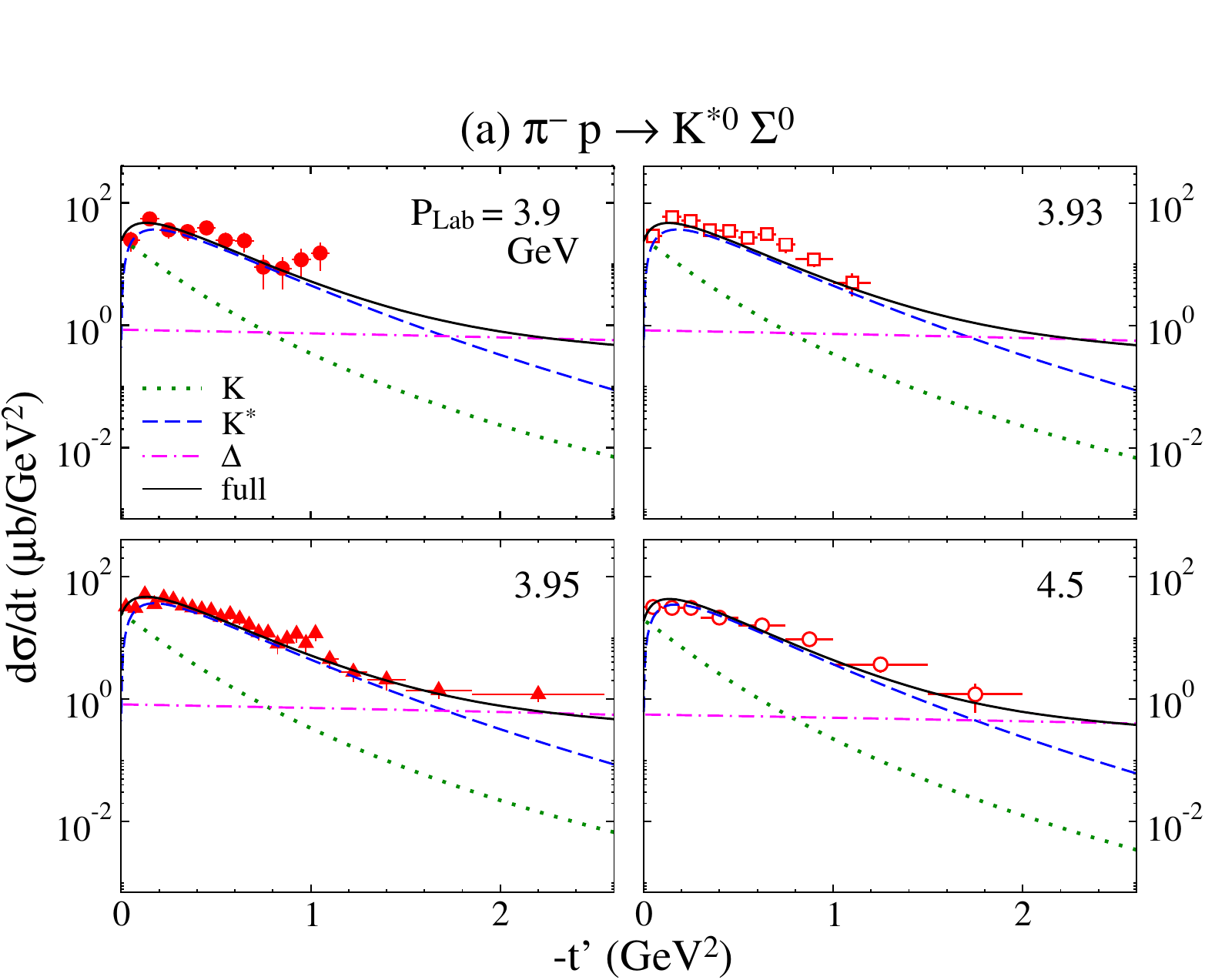} \hspace{1em}
\includegraphics[width=\columnwidth]{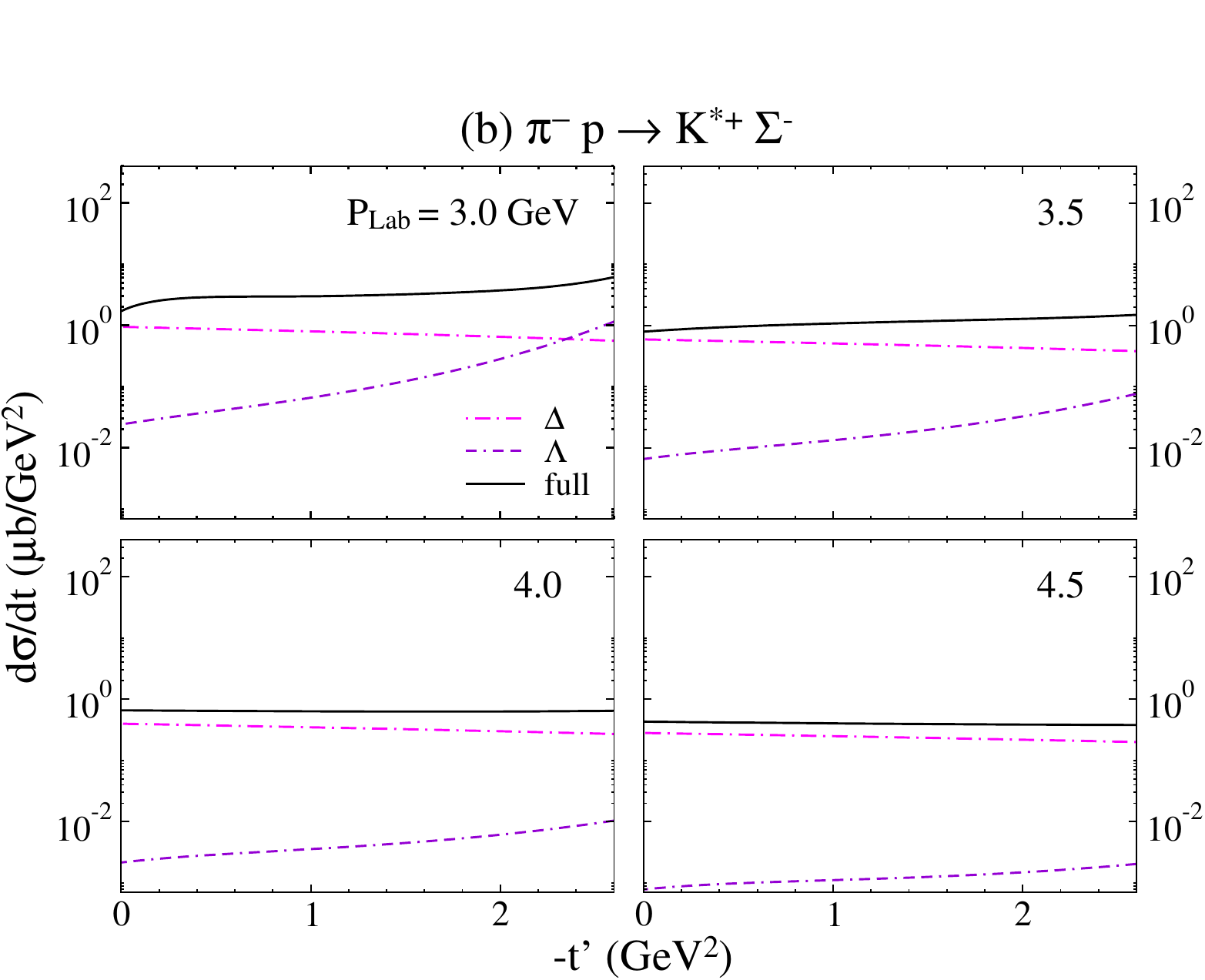}
\caption{Differential cross sections as functions of $-t'$ for the (a) $\pi^- p \to
K^{*0} \Sigma^0$ and (b) $\pi^- p \to K^{*+}\Sigma^-$ reactions at four fixed beam
energies.
The experimental data in panel (a) are taken from Refs.~\cite{Abramovich:1972rq}
(circles), \cite{Yaffe:1973ex} (open squares), \cite{CCMS:1980mch} (triangles), and
\cite{Crennell:1972km} (open circles).}
\label{FIG06}
\end{figure*}
\begin{figure}[ht]
\centering
\includegraphics[width=4.2cm]{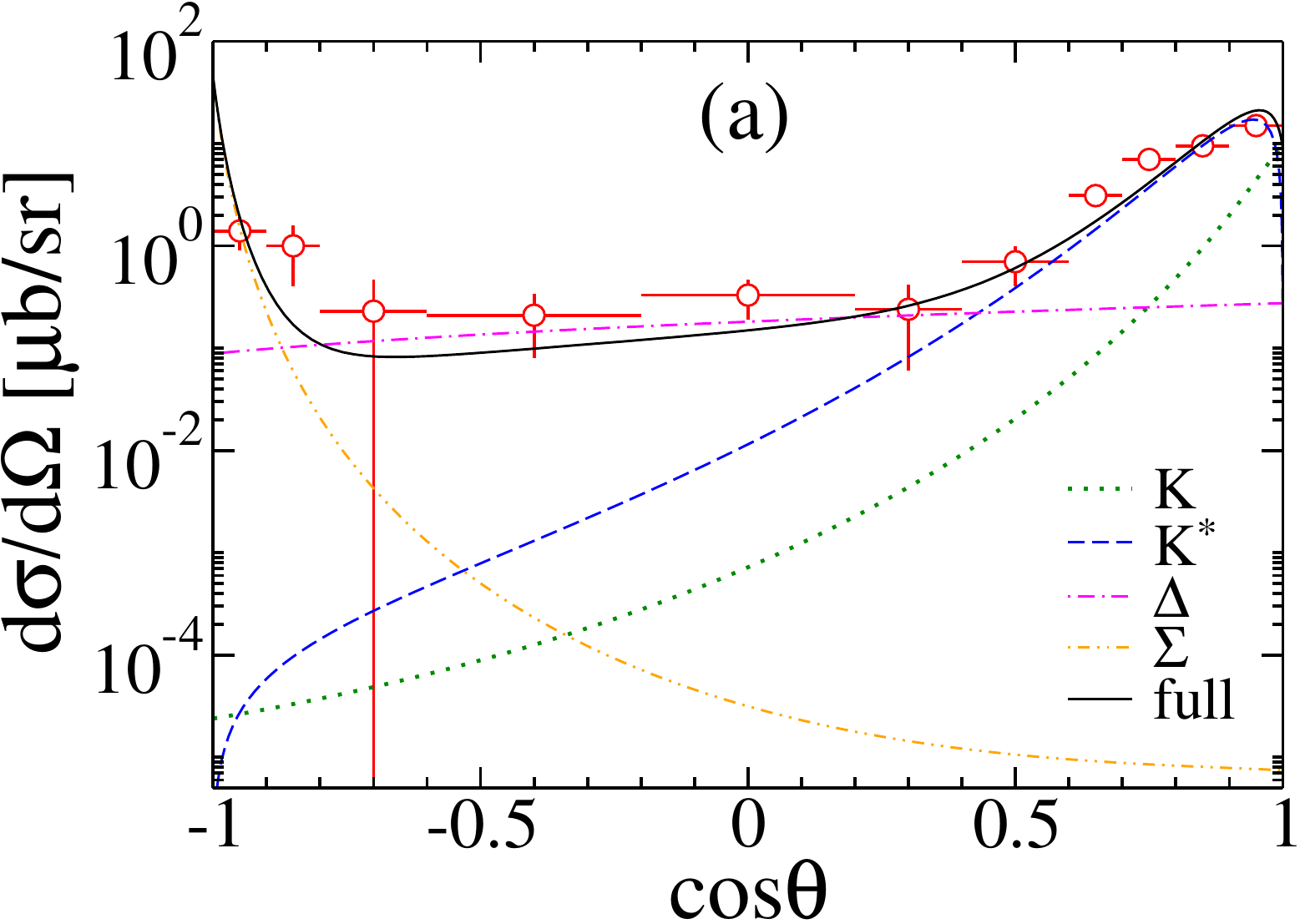}
\includegraphics[width=4.2cm]{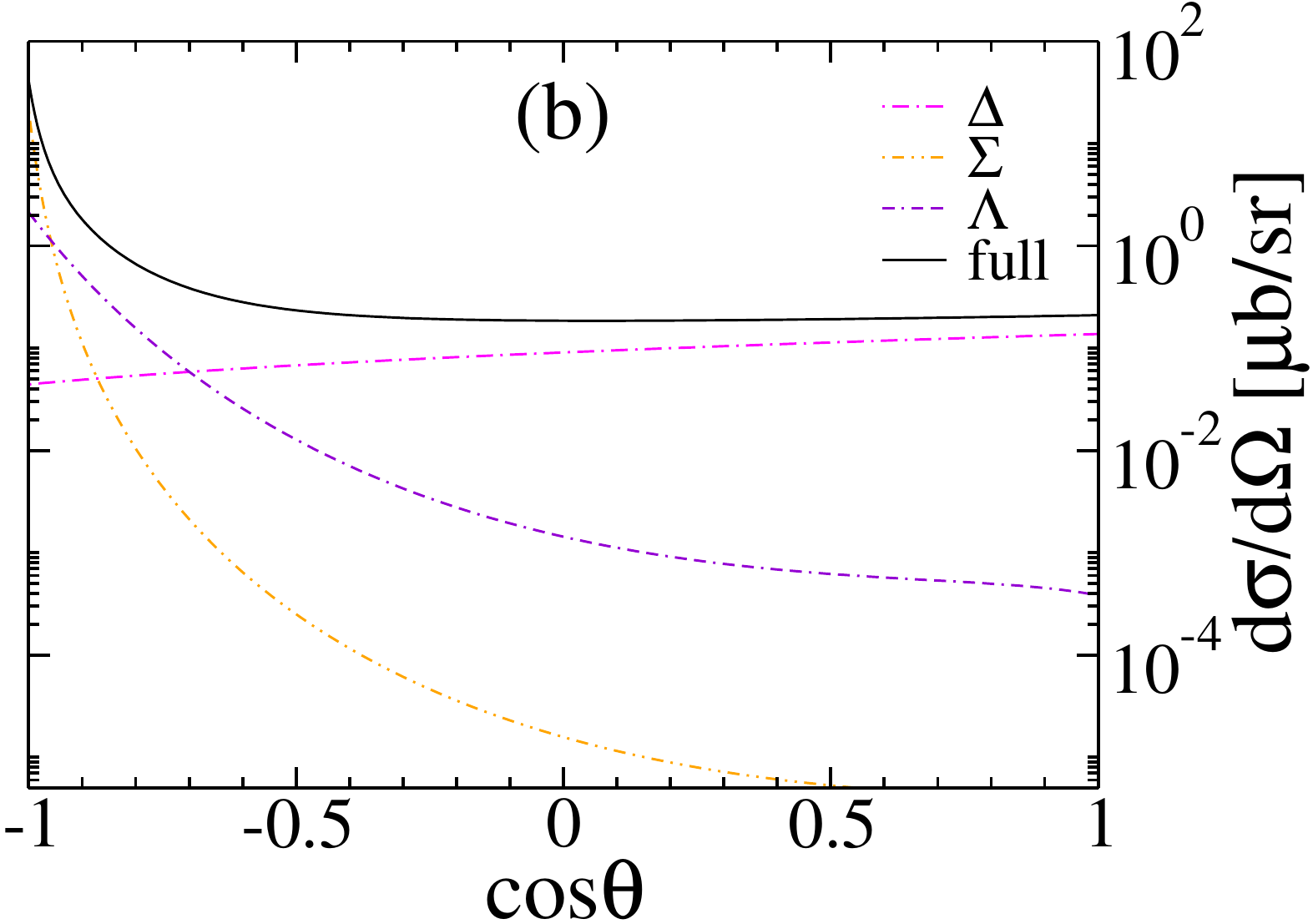}
\caption{Differential cross sections as functions of $\cos\theta$ for the (a) $\pi^-
p \to K^{*0} \Sigma^0$ and (b) $\pi^- p \to K^{*+} \Sigma^-$ reactions at
$P_{\mathrm{Lab}}$ = 4.5 GeV.
The experimental data in panel (a) are taken from Ref.~\cite{Crennell:1972km}.}
\label{FIG07}
\end{figure}
The results in Fig.~\ref{FIG04} reveal sizable discrepancies between the theoretical
predictions and the experimental data~\cite{Dahl:1967pg,Miller:1965,Crennell:1972km,
CCMS:1980ysu} for $P_{\mathrm{Lab}} \lesssim 3$ GeV,
suggesting that additional mechanisms such as the $s$-channel $N^*$ and $\Delta^*$
resonances (see Fig.~\ref{FIG03}) may play a crucial role in this energy region.
Figure~\ref{FIG05} displays the full results for the total cross sections including
the $N^*$ and $\Delta^*$ resonance contributions to the Born terms as functions of
the c.m. energy $W$.
For the $s$-channel $N^*$ and $\Delta^*$ exchanges, the isospin factors are $I_{N^*}$ =
$-\sqrt{2}$ and $I_{\Delta^*}$ = $\sqrt{2}/3$  for the (a) $K^{*0}\Sigma^0$ channel,
and $I_{N^*}$ = $2$ and $I_{\Delta^*}$ = $1/3$ for the (b) $K^{*+}\Sigma^-$ channel,
yielding $\sigma(K^{*0}\Sigma^0)/\sigma(K^{*+}\Sigma^-)$ = $0.5$ and $2$ for the $N^*$
and $\Delta^*$ exchanges, respectively.
The much larger discrepancies between the Born contributions and the experimental
data for the (a) $K^{*0}\Sigma^0$ than (b) $K^{*+}\Sigma^-$ channel therefore
imply that the $\Delta^*$ resonances may play a more important role than the $N^*$
ones at $P_{\mathrm{Lab}} \lesssim 3$ GeV.
Indeed, we find that the $\Delta(2150)1/2^-$ resonance is chiefly responsible for the
near-threshold enhancement in both reactions.
This is expected from its much larger branching ratio
$\mathrm{Br}_{\Delta(2150) \to K^* \Sigma}$ compared with those of the other resonances
(see Table~\ref{TAB:3}).
The peak structure around 2.1 GeV arises from the $\Delta(2150) 1/2^-$
resonance and is prominent in both reactions.
The contributions from other $N^*$ and $\Delta^*$ resonances are small but
non-negligible.
Their combined effect produces a second peak around 2.4 GeV and is visible only in
the (a) $K^{*0}\Sigma^0$ channel.

In Fig.~\ref{FIG06}, the differential cross sections are plotted as functions of
$-t' = -t + t_{min}$.
The experimental data are available only for the (a) $K^{*0}\Sigma^0$ channel for
$P_{\mathrm{Lab}}$ = 3.9--4.5 GeV and they help to constrain the $t$-channel $K$- and
$K^*$-Reggeon exchanges at forward angles and low momentum transfer.
As shown in Fig.~\ref{FIG06}(a), the overall $t$ dependence is governed mainly by 
$K^*$-Reggeon exchange, although distinct behavior appears at very forward angles.
For $K$-Reggeon exchange, the cross section increases monotonically, whereas the
$K^*$-Reggeon exchange suppresses the cross section in the very forward region.
The coherent sum of the two exchanges provides a reasonable description of the
experimental data~\cite{Abramovich:1972rq,Yaffe:1973ex,CCMS:1980mch,
Crennell:1972km}.
The effect of the $\Sigma$-Reggeon exchange becomes noticeable at $-t' \gtrsim$ 1.5
GeV$^2$.
In Fig.~\ref{FIG06}(b), we present the predictions of the differential cross sections
for the (b) $K^{*+}\Sigma^-$ channel for $P_{\mathrm{Lab}}$ = 3.0--4.5 GeV, where the
process is dominated by the $s$-channel $\Delta$ exchange in the considered region.
As a result, the differential cross section is nearly flat over the kinematic range
shown.
Conversely, the differential cross section $d\sigma/du$ would be useful for
investigating the effects of the $\Sigma$- and $\Lambda$-Reggeon exchanges.

In Fig.~\ref{FIG07}, the differential cross sections are plotted as functions of
$\cos\theta$ at $P_{\mathrm{Lab}} = 4.5$ GeV where $\theta$ is the scattering angle of
the produced $K^*$ meson in the c.m. frame.
For the (a) $K^{*0}\Sigma^0$ channel, the roles of the $K$- and $K^*$-Reggeon
exchanges are again confirmed by comparison with the available experimental
data~\cite{Crennell:1972km}.
The $\Delta$ exchange makes the angular distribution relatively flat at
intermediate scattering angles.
Backward peaking is also observed and is described by the $\Sigma$-Reggeon exchange.
For the (b) $K^{*+}\Sigma^-$ channel, the $\Sigma$- and $\Lambda$-Reggeon exchanges
provide dominant contributions at backward angles, leading to a pronounced
backward-peaking behavior.
At $\cos\theta \gtrsim -0.5$, the reaction is mainly governed by the $\Delta$
exchange, resulting in a relatively flat angular distribution.

\begin{figure*}[ht]
\centering
\includegraphics[width=8.5cm]{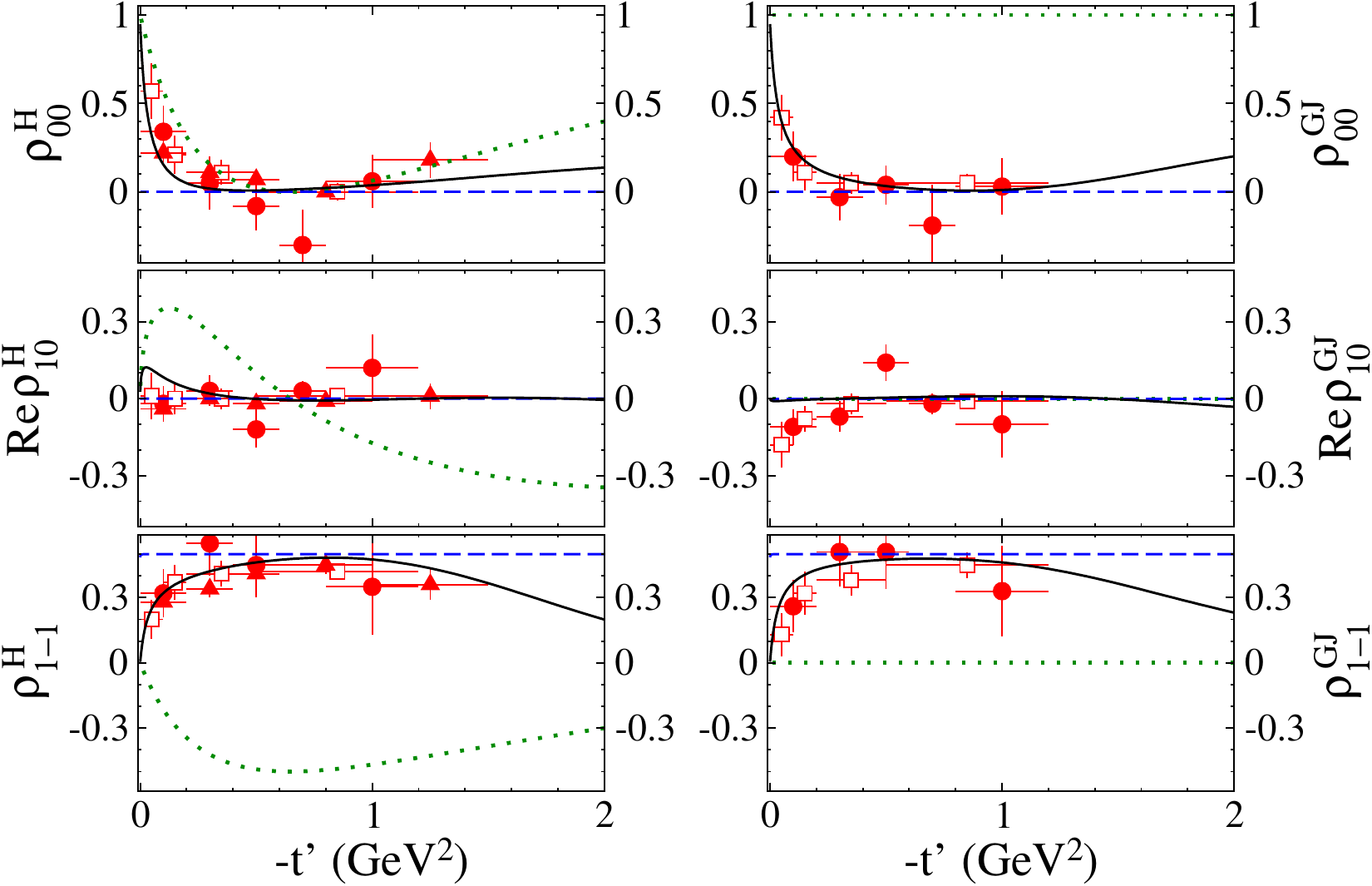} \hspace{1em}
\includegraphics[width=8.5cm]{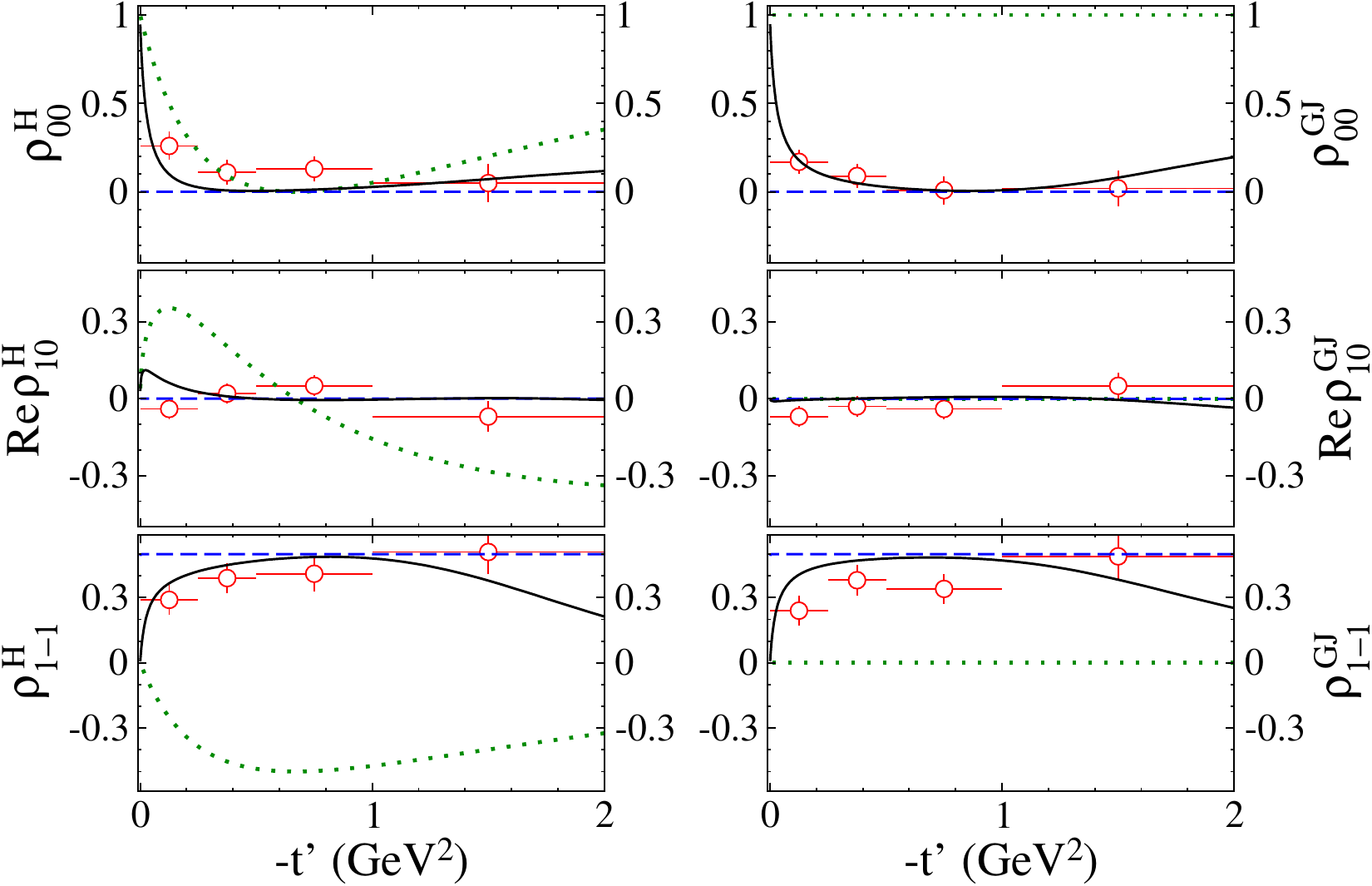}
\caption{SDMEs $\rho_{00}$, $\mathrm{Re}\rho_{10}$, and $\rho_{1-1}$ in the helicity
and GJ frames for the $\pi^- p \to K^{*0}\Sigma^0$ reaction as functions of $-t'$ at
$P_{\mathrm{Lab}}$ = 3.93 and 4.5 GeV in the left and right panels, respectively.
The green dotted and blue dashed curves denote the $K$- and $K^*$-Reggeon
contributions, respectively, while the black solid curves represent the full results.
Experimental data are taken from Refs.~\cite{Abramovich:1972rq} (circles),
\cite{Yaffe:1973ex} (open squares), and \cite{CCMS:1980mch} (triangles) for the
left panels, and from Ref.~\cite{Crennell:1972km} (open circles) for the right
panels.
The beam momenta in Refs.~\cite{Abramovich:1972rq} and \cite{CCMS:1980mch} are
$P_{\mathrm{Lab}}$ = 3.9 GeV and 3.95 GeV, respectively.}
\label{FIG08}
\end{figure*}
\begin{figure}[ht]
\centering
\includegraphics[width=8.5cm]{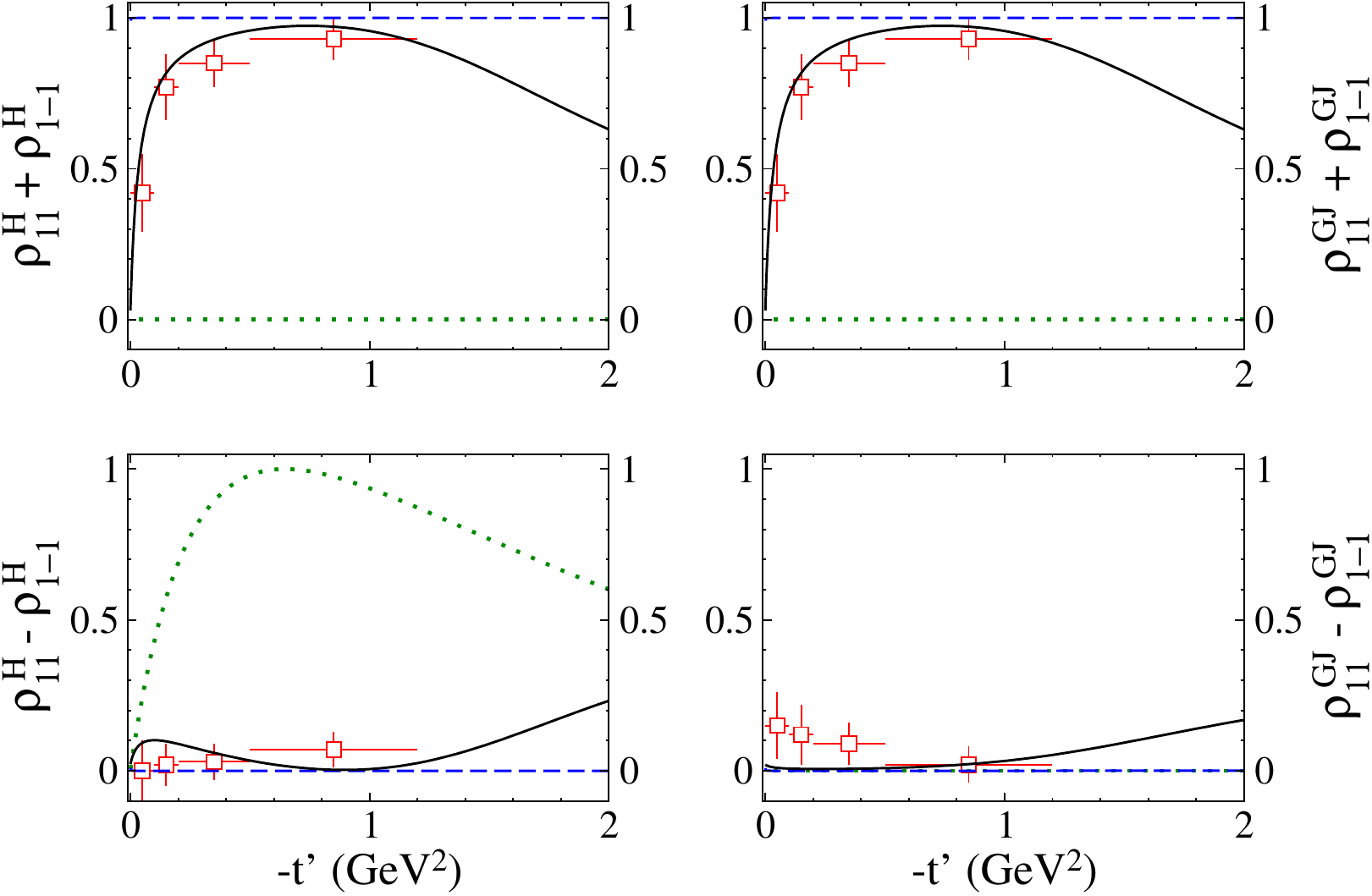}
\caption{SDMEs $\rho_{11} + \rho_{1-1}$ and $\rho_{11} - \rho_{1-1}$ in the helicity and
GJ frames for the $\pi^- p \to K^{*0}\Sigma^0$ reaction as functions of $-t'$ at
$P_{\mathrm{Lab}}$ = 3.93 GeV.
Curve notations are the same as in Fig.~\ref{FIG08}.
Experimental data are taken from Ref.~\cite{Yaffe:1973ex}.}
\label{FIG09}
\end{figure}
\begin{figure}[ht]
\centering
\includegraphics[width=8.5cm]{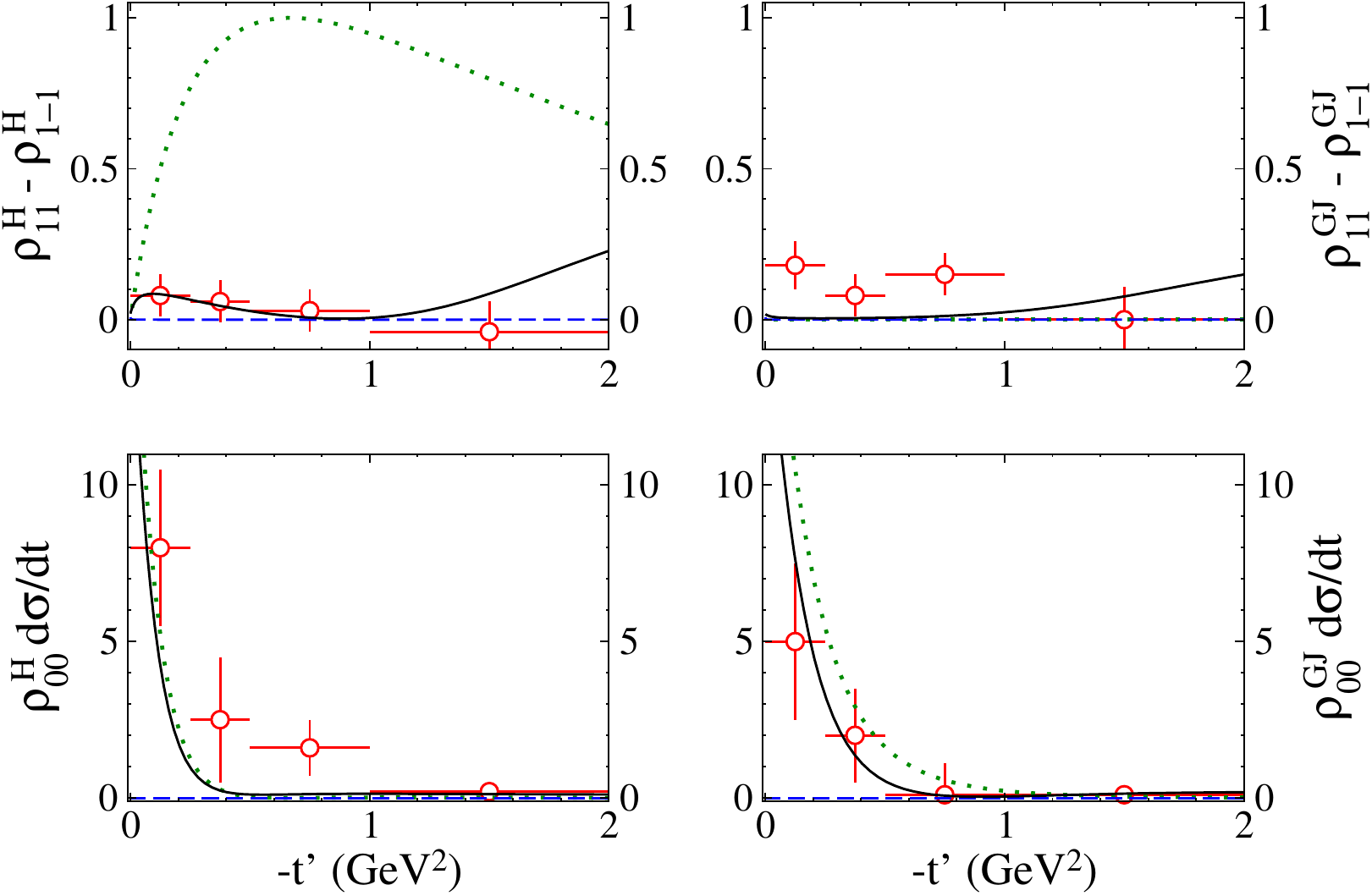}
\caption{SDMEs $\rho_{11} - \rho_{1-1}$ and $\rho_{00}\, d\sigma/dt$ in the helicity and
GJ frames for the $\pi^- p \to K^{*0}\Sigma^0$ reaction as functions of $-t'$ at
$P_{\mathrm{Lab}}$ = 4.5 GeV.
Curve notations are the same as in Fig.~\ref{FIG08}.
Experimental data are taken from Ref.~\cite{Crennell:1972km}.}
\label{FIG10}
\end{figure}
It is worthwhile to examine the SDMEs, which carry information on the helicity
structure of the reaction amplitude and thus help to clarify the underlying
production mechanism.
In this work, we consider the case where the produced $K^*$ vector meson is only
polarized.
The case where the recoil $\Sigma$ baryon is also polarized is discussed in
Ref.~\cite{Kim:2017hhm}.

For definiteness, we consider the decay mode $K^* \to K\pi$ in the $\pi^- p \to
K^* \Sigma$ reaction.
When analyzing the momentum distribution of the final-state kaon in the $K^*$ rest
frame, one must specify a quantization axis.
One choice is to take the axis opposite to the momentum of the outgoing $\Sigma$
hyperon in the $K^*$ decay.
Another possible definition is to align the axis with the momentum of the incoming
meson.
According to the conventions introduced in Refs.~\cite{Crennell:1972km,
Schilling:1969um}, the first definition corresponds to the helicity (H) frame,
whereas the second is called the Gottfried–Jackson (GJ) frame.
The helicity frame is typically employed to examine $s$-channel helicity
conservation, while the GJ frame is more appropriate for studying the $t$-channel
exchange mechanism.

\begin{figure*}[ht]
\centering
\includegraphics[width=7.0cm]{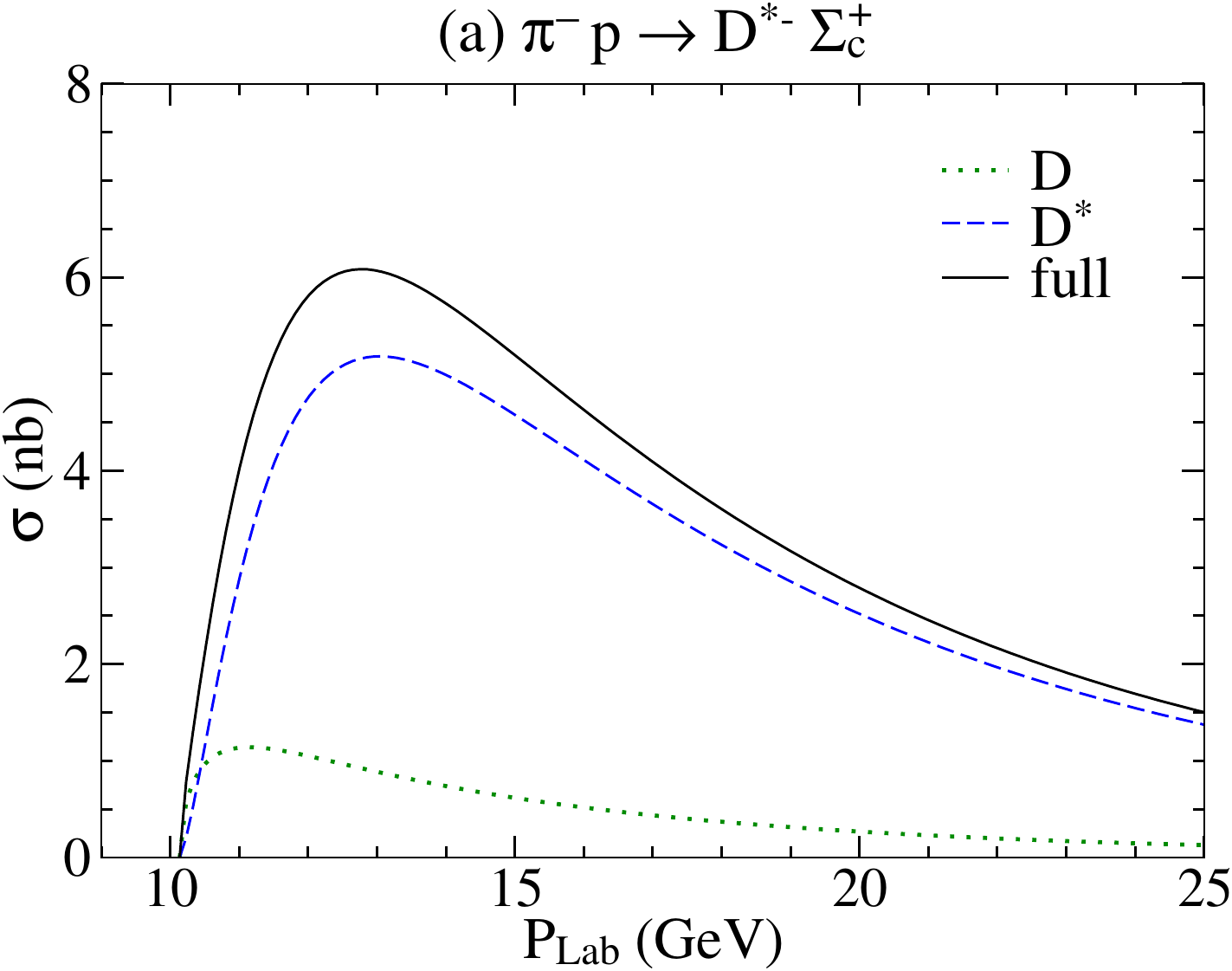} \hspace{1em}
\includegraphics[width=7.4cm]{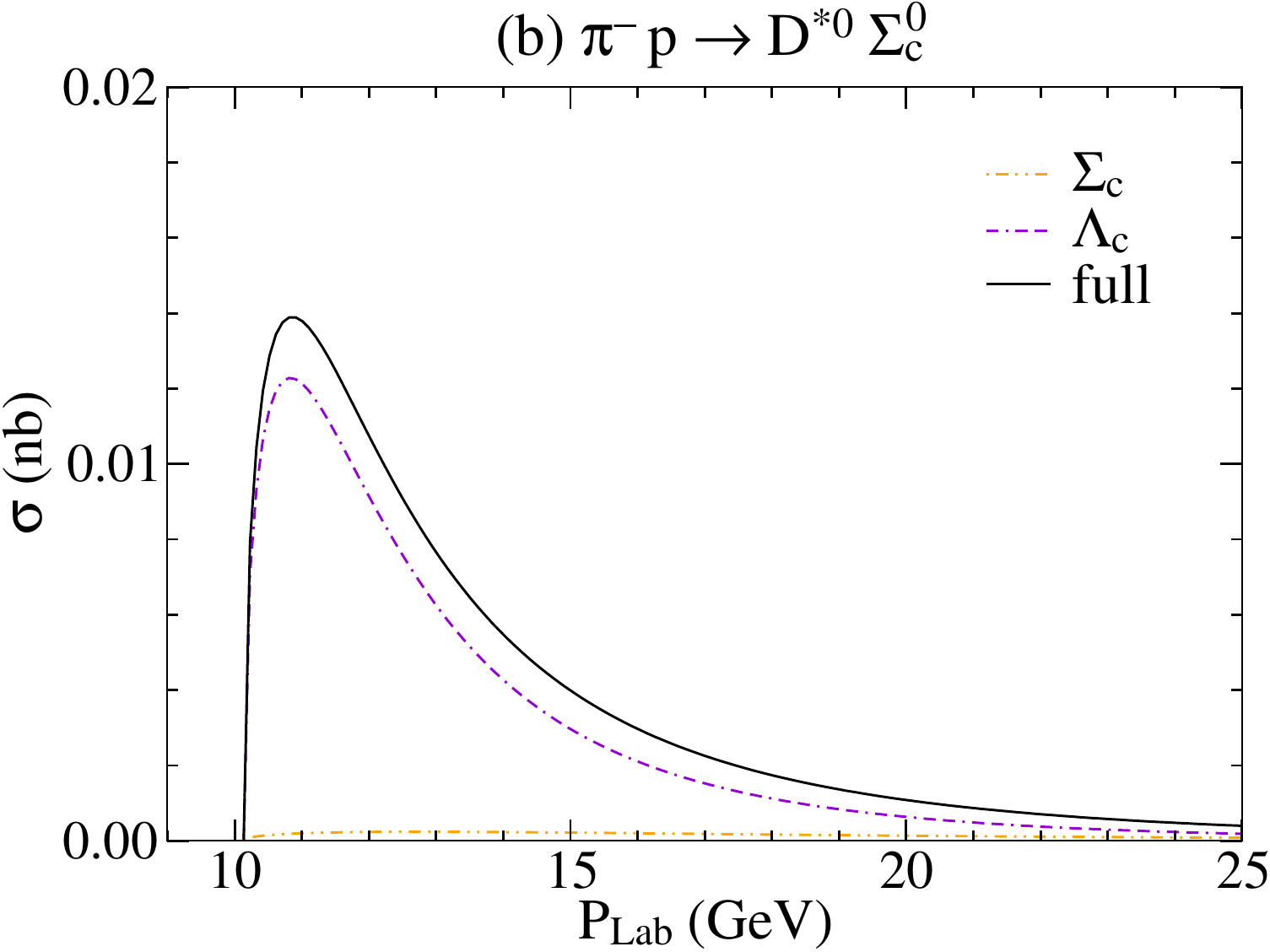}
\caption{Total cross sections for the (a) $\pi^- p \to D^{*-} \Sigma_c^+$ and (b)
$\pi^- p \to D^{*0} \Sigma_c^0$ reactions as functions of $P_{\rm Lab}$.}
\label{FIG11}
\end{figure*}
\begin{figure}[ht]
\centering
\includegraphics[width=7.0cm]{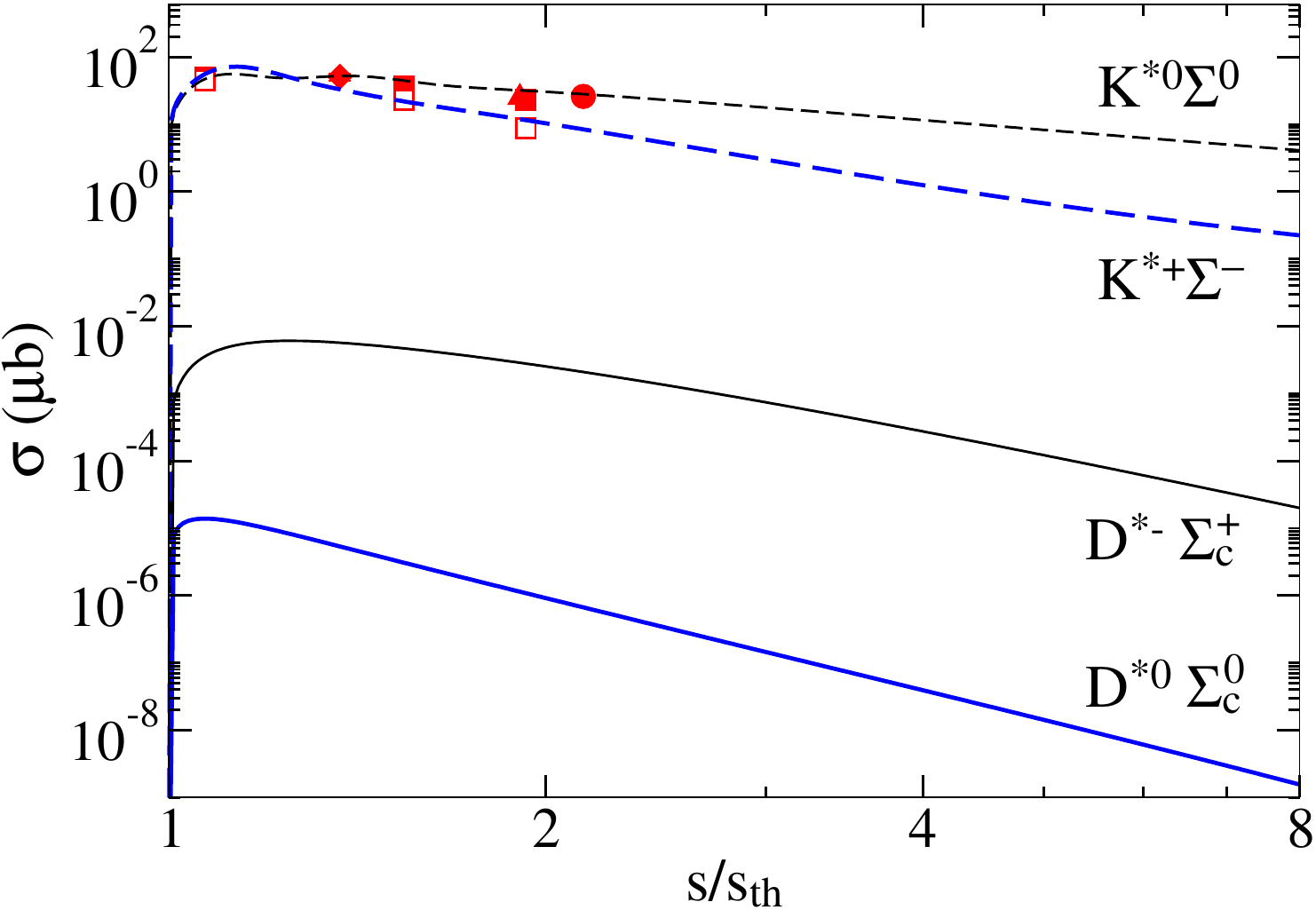}
\caption{Total cross sections for the open-strangeness $\pi^- p \to$ ($K^{*0}
\Sigma^0$, $K^{*+} \Sigma^-$) and open-charm $\pi^- p \to$ ($D^{*-} \Sigma_c^+$,
$D^{*0} \Sigma_c^0$) reactions as functions of $s/s_{\mathrm{th}}$.}
\label{FIG12}
\end{figure}
The decay probabilities are expressed in terms of the SDMEs $\rho_{\lambda \lambda'}$,
where $\lambda = \lambda_V$.
These SDMEs are constructed from the reaction amplitudes given in
Eqs.~(\ref{eq:RegAmpl1}), (\ref{eq:RegAmpl2}), (\ref{eq:RegAmpl3}),
(\ref{eq:RegAmpl4}), and the related expressions:
\begin{align}
\rho_{\lambda\lambda'} = \frac{1}{{\mathcal N}^2}
\sum\limits_{s_f = \pm\frac12,\, s_i = \pm\frac12}
T_{\lambda,s_f;s_i}\, T^*_{\lambda',s_f;s_i},
\label{eq:SDME}
\end{align}
where the normalization factor is given by
\begin{align}
{\mathcal N}^2 = \sum_{\lambda_,s_f,s_i} |T_{\lambda,s_f ; s_i}|^2.
\label{eq:SDME-NF}
\end{align}
The Hermitian conditions,
\begin{align}
\rho_{1-1} = \rho_{-11}, \,\,\,
\rho_{10} = \rho_{01}, \,\,\,
\rho_{-10} = \rho_{0-1},
\label{eq:SDME}
\end{align}
are imposed in our calculations.
In addition, the normalization condition $\rho_{00}+\rho_{11}+\rho_{-1-1}=1$ and the
symmetry relations,
\begin{align}
&\rho_{11} = \rho_{-1-1},\,\,\, \rho_{\pm 10} = \rho_{0 \pm 1},
\cr
&\rho_{1-1} = \rho_{-11},\,\,\, \rho_{\pm 10} = -\rho_{0 \mp 1},
\end{align}
are satisfied.

Experimental data for the $t$-dependent SDMEs, $\rho_{00}$, $\mathrm{Re}\rho_{10}$,
and $\rho_{1-1}$, are available for the $K^{*0}\Sigma^0$ reaction in both the helicity
and GJ frames~\cite{Abramovich:1972rq,Yaffe:1973ex,CCMS:1980mch,Crennell:1972km}.
We compare our results with these data in Fig.~\ref{FIG08}.
The results at $P_{\mathrm{Lab}}$ = 3.93 and 4.5 GeV are shown in the left and right
panels, respectively, and both panels exhibit similar features.
In principle, for the vector-meson exchange, the matrix elements
$\rho_{\lambda\lambda'}$ with $|\lambda| = |\lambda'| = 1$ are specifically enhanced
because of the spin structure $\epsilon^{\mu\nu\alpha\beta}\varepsilon^*_\mu(\lambda_V)
k_{2\alpha} k_{1\beta}$ of the amplitude in Eq.~(\ref{eq:Ampl1}).
In the vector-meson rest frame, where $\bm{k}_2=(M_V,0,0,0)$ and $\bm{k}_1$
corresponds to the three-momentum of the incoming pion, this factor becomes
proportional to the vector product $\bm{\varepsilon}^* (\lambda_V) \times \bm{k}_1$.
In the helicity frame and at small momentum transfers, $\bm{k}_1$ has a large
$z$ component and only a small transverse component.
Consequently, $\bm{\varepsilon}^*(\lambda_V) \times \bm{k}_1 \simeq
i\lambda_V\,\bm{\varepsilon}^*(\lambda_V)|\bm{k}_1|$, which leads to a pronounced
enhancement of $\rho_{|\lambda|=1,\,|\lambda'|=1}$.
In the GJ frame, $\bm{k}_1$ is aligned with the quantization axis; therefore,
$\rho_{\lambda\lambda'}$ vanishes whenever either $\lambda=0$ or $\lambda'=0$.
Accordingly, for the $K^*$-Reggeon exchange, only $\rho_{1-1}$ is found to be
significantly enhanced, while $\rho_{00}$ and $\mathrm{Re}\rho_{10}$ remain close to
zero.
For the $K$-Reggeon exchange, however, the situation is totally different.
In this case, the scattering amplitude is proportional to the scalar product
$\bm{\varepsilon}^*(\lambda_V) \cdot \bm{k}_1$ in Eq.~(\ref{eq:Ampl1}), which
strongly enhances $\rho_{00}$ in the GJ frame.
As a result, $\rho_{00}^{\mathrm{GJ}}$  = 1 while all other
$\rho_{\lambda\lambda'}^{\mathrm{GJ}}$ vanish~\cite{Kim:2017hhm,Kim:2026tgd}.
The results in Fig.~\ref{FIG08} clearly show that the coherent sum of the $K$- and
$K^*$-Reggeon exchanges is in excellent agreement with the experimental data.

Figure~\ref{FIG09} shows the $t$-dependent SDMEs, $\rho_{11} + \rho_{1-1}$ and 
$\rho_{11} - \rho_{1-1}$, for the $K^{*0} \Sigma^0$ reaction in both the helicity
and GJ frames at $P_{\mathrm{Lab}}$ = 3.93 GeV.
Note that, by definition, for the $K^*$-Reggeon exchange, $\rho_{11}^{\mathrm{GJ}}$ is
consistent with 0.5 and $\rho_{1-1}$ is smaller than $\rho_{11}$ in both frames.
In our case, however, $\rho_{1-1}$ is enhanced to nearly the same magnitude as
$\rho_{11}$.
Consequently, $\rho_{11} + \rho_{1-1}$ is close to 1, while $\rho_{11} - \rho_{1-1}$
is approximately zero.
For the $K$-Reggeon exchange, one finds that $\rho_{11}^{\mathrm{H}}$ =
$- \rho_{1-1}^{\mathrm{H}}$, which leads to $\rho_{11}^{\mathrm{H}} +
\rho_{1-1}^{\mathrm{H}}$ = 0.
In addition, as mentioned earlier, $\rho_{11}^{\mathrm{GJ}}$ = $\rho_{1-1}^{\mathrm{GJ}}$
= 0, so that $\rho_{11}^{\mathrm{GJ}} \pm \rho_{1-1}^{\mathrm{GJ}}$ = 0.
The full results in Fig.~\ref{FIG09} reproduce the experimental data~\cite{
Yaffe:1973ex} quite well.
The decrease of $\rho_{11} + \rho_{1-1}$ and the increase of $\rho_{11} - \rho_{1-1}$
at $-t' \gtrsim$ 1 GeV$^2$ in both frames originate from the $\Delta$-baryon
contribution.
In Fig.~\ref{FIG10}, we present the $t$-dependent SDMEs, $\rho_{11} - \rho_{1-1}$ and
$\rho_{00}\, d\sigma/dt$, for the $K^{*0} \Sigma^0$ reaction in both the helicity
and GJ frames at $P_{\mathrm{Lab}}$ = 4.5 GeV~\cite{Crennell:1972km}.
The results for $\rho_{11} - \rho_{1-1}$ are very similar to those shown in the
bottom panels of Fig.~\ref{FIG09}.
The shapes of $\rho_{00}\, d\sigma/dt$ are consistent with those of $\rho_{00}$
shown in the top panels of Fig.~\ref{FIG08}.

We now turn to charm production in $\pi^- p \to D^* \Sigma_c$, using the framework
described in Secs.~\ref{Sec:II-2} and \ref{Sec:II-3}.
The cutoff masses are taken from the strangeness-sector analysis, allowing us to make
predictions for the charm sector within the same parameter set.
Figure~\ref{FIG11} shows the predicted total cross sections for the (a) $\pi^- p
\to D^{*-} \Sigma_c^+$ and (b) $\pi^- p \to D^{*0} \Sigma_c^0$ reactions as functions
of $P_{\mathrm{Lab}}$.
For the (a) $D^{*-} \Sigma_c^+$ channel, $D^*$-Reggeon exchange is the most
significant, while $D$-Reggeon exchange also plays a non-negligible role near
threshold.
For the (b) $D^{*0} \Sigma_c^0$ channel, $\Lambda_c$-Reggeon exchange dominates
the process.
Because the $t$-channel contribution is absent, the total cross section is about two
orders of magnitude smaller than that for the  (a) $D^{*-} \Sigma_c^+$ channel.
In both reactions, the $\Sigma_c$-Reggeon contribution is strongly suppressed.

We compare the total cross sections for the open-charm reactions $\pi^- p \to$
($D^{*-} \Sigma_c^+$, $D^{*0} \Sigma_c^0$) with those for the open-strangeness reactions
$\pi^- p \to$ ($K^{*0} \Sigma^0$, $K^{*+} \Sigma^-$) in Fig.~\ref{FIG12}, plotted as
functions of $s/s_{\mathrm{th}}$, where $s_{\mathrm{th}}$ is the threshold energy of the
corresponding reaction.
Near threshold, the magnitudes of the two strangeness reactions are similar; however,
as the energy increases, the cross section for the $K^{*+} \Sigma^-$ reaction
decreases more rapidly than that for the $K^{*0} \Sigma^0$ reaction.
This behavior originates from the smaller Regge intercept of the $\Lambda$,
$\alpha_{\Lambda}(0)$, compared to that of the $K^*$, $\alpha_{K^*}(0)$.
The energy dependence of the cross sections is governed by the Regge
intercept~[see Eqs.~(\ref{eq:Asym:dsdt}) and (\ref{eq:Asym:dsdu})].
Depending on $s/s_{\mathrm{th}}$, the $D^{*-} \Sigma_c^+$ cross section is suppressed by
about $4$--$5$ orders of magnitude relative to $K^{*0}\Sigma^0$, while the $D^{*0}
\Sigma_c^0$ cross section is suppressed by about $7$--$8$ orders of magnitude relative
to $K^{*+} \Sigma^-$.
These suppressions can be mainly attributed to the larger energy-scale parameter in
the charm sector, $s_{D (D^*)}^{\pi N : D^* \Sigma_c} > s_{K (K^*)}^{\pi N : K^* \Sigma}$.
The cross sections for charm production in the $D^{*-} \Sigma_c^+$ and $D^{*0}
\Sigma_c^0$ channels are expressed in units of nb and pb, respectively.
Consequently, future high-precision accelerator facilities will be necessary to
probe such small cross sections, particularly for the latter reaction.

\section{Summary and Conclusion}
\label{Sec:IV}

In this work, we investigated the $\pi^- p \to K^* \Sigma$ reaction within a hybrid
Regge framework.
In the $K^{*0} \Sigma^0$ channel, the reaction is governed mainly by $t$-channel $K$-
and $K^*$-Reggeon exchanges, while the $\Delta$ and $\Sigma$-Reggeon exchanges are
important in the intermediate- and backward-angle regions, respectively.
By contrast, in the $K^{*+} \Sigma^-$ channel, the $t$-channel contribution is absent,
and the $u$-channel $\Lambda$-Reggeon exchange becomes essential, producing a
backward-peaked angular distribution.
We also included several $s$-channel $N^*$ and $\Delta^*$ resonances to describe the
near-threshold structure and found that the $\Delta (2150)1/2^-$ resonance provides
the dominant resonance contribution because of its strong coupling to the $K^*
\Sigma$ channel.

Polarization observables offer valuable insight into the underlying reaction
mechanism, and we examined the SDMEs in both the helicity and GJ frames in detail.
The roles of the $K$- and $K^*$-Reggeon exchanges are more clearly identified in the
$K^{*0} \Sigma^0$ channel, leading to excellent agreement with the available
experimental data.
Future measurements of differential cross sections and SDMEs near threshold ($W
\lesssim$ 2.5 GeV) would further strengthen our conclusions on the contributions of
$s$-channel baryon resonances.

While we extract the $R \to K^* \Sigma$ couplings from quark-model estimates~\cite{
Capstick:1998uh}, alternative interpretations exist in the literature.
For example, Refs.~\cite{Khemchandani:2011et,Lin:2018kcc} interpret the $N(2080)$
resonance as a $K^* \Sigma$ bound state and as a strange partner of the heavy
pentaquark $P_c$ states~\cite{LHCb:2015yax,LHCb:2019kea,LHCb:2021chn}, with
$J^P = 3/2^-$.
Its contribution to the $\pi^- p \to K^* \Sigma$ reaction has also been discussed in
Ref.~\cite{Wang:2024xvq}.
Moreover, Ref.~\cite{Khemchandani:2011et} suggests that the $\Delta(1900)1/2^-$
resonance couples strongly to the $K^* \Sigma$ channel.
In addition, subthreshold resonances such as $N(1535)1/2^-$, $N(1650)1/2^-$,
$N(1895)1/2^-$, and $\Delta(1600)3/2^+$ may also contribute~\cite{
Khemchandani:2013nma}.
These effects are not included in the present work and deserve further study.

Using the same framework, with Regge trajectories and energy-scale parameters fixed
by a QGSM-motivated prescription, we also studied the $\pi^- p \to D^* \Sigma_c$
reaction.
The $D^{*-} \Sigma_c^+$ and $D^{*0} \Sigma_c^0$ cross sections are found to be strongly
suppressed relative to their strangeness counterparts, by about $4$--$5$ and 
$7$--$8$ orders of magnitude, respectively, depending on $s/s_{\mathrm{th}}$.
This suppression is mainly caused by the larger energy-scale parameters in the charm
sector.
Our results for both strangeness and charm production, including total and
differential cross sections as well as SDMEs, should be useful for future
measurements at J-PARC~\cite{Aoki:2021cqa} and AMBER~\cite{Adams:2018pwt}.

In our previous work, the $\pi^- p \to K^{*+} \Lambda$ and $\pi^- p \to D^{*-}
\Lambda_c^+$ reactions were studied within the same framework, but only the Born
terms were included~\cite{Kim:2015ita,Kim:2017hhm}.
Extending the $\pi^- p \to K^{*+} \Lambda$ analysis to include $s$-channel $N^*$
resonances would be particularly valuable, since this channel acts as an isospin
filter for $N^*$ states.
The present study can also be extended to kaon-induced reactions, such as $K^- p
\to \phi \Sigma^0$, $D_s^{*-} \Sigma_c^+$, and $J/\psi \Sigma^0$~\cite{Kim:2026tgd}.
Related studies are under way.

\section*{Acknowledgments}

This work was supported by the National Research Foundation of Korea (NRF) grant
funded by the Korea government (Nos. RS-2021-NR060129 and RS-2026-25468435).



\end{document}